\documentclass{jpp}

\usepackage{graphicx}
\usepackage{color}
\usepackage{epstopdf, epsfig}
\usepackage{hyperref}

\usepackage{bm}
\usepackage{amsmath,amssymb}

\newcommand{\rot}{\nabla\times}

\newcommand{\EB}{\bm{E} \times \bm{B}}
\newcommand{\imag}{\mathrm{i}}
\newcommand{\diff}{\mathrm{d}}
\newcommand{\odf}[2]{\frac{\diff #1}{\diff #2}}
\newcommand{\pdf}[2]{\frac{\partial #1}{\partial #2}}

\newcommand{\revise}[1]{\textcolor{black}{#1}}
\newcommand{\revproof}[1]{\textcolor{black}{#1}}

\newcommand{\mathsfit}[1]{\text{\sffamily\slshape #1}}

\shorttitle{Random forcing for gyrokinetic simulations}
\shortauthor{R. Numata}

\title{Random forcing with a constant power input for two-dimensional gyrokinetic simulations}

\author{Ryusuke Numata
  \corresp{\email{ryusuke.numata@gmail.com}}}

\affiliation{Graduate School of Simulation Studies, University of Hyogo,
7-1-28 Minatojima Minami-machi, Chuo-ku, Kobe, Hyogo 650-0047, Japan}

\begin{document}

\maketitle

\begin{abstract}
  A method of random forcing with a constant power input for
  two-dimensional gyrokinetic turbulence simulations is developed for
  the study of stationary plasma turbulence. The property that the
  forcing term injects the energy at a constant rate enables
  \revproof{turbulence to be set up}
  in the desired range and \revproof{energy dissipation channels to be
  assessed quantitatively}
  in a statistically steady state. Using the
  developed method, \revproof{turbulence is demonstrated in the
  large-scale fluid and small-scale kinetic regimes}, 
  where the theoretically predicted scaling laws are
  \revproof{reproduced successfully}.
\end{abstract}

\section{Introduction}
\label{sec:introduction}

Turbulence and magnetic reconnection are key fundamental nonlinear
processes in weakly collisional plasmas governing energy conversion and
transport. Astrophysical plasmas, such as solar winds, accretion \revproof{discs},
the interstellar medium, as well as fusion plasmas are mostly
collisionless and in a turbulent state~\citep{Biskamp_03}. Plasmas tend
to form current sheets where magnetic reconnection occurs to release the
magnetic energy and convert it into heat~\citep{Biskamp_00}.
\revproof{Turbulence} and magnetic reconnection \revproof{often}
coincide. Solar winds are the most typical examples of such a situation
\citep[e.g.][]{RetinoSundkvistVaivads_07} and provide rich observational
information about these processes.

According to the conventional wisdom of turbulence (see
e.g. \citet{Frisch_95} for neutral fluids \revproof{and}
\citet{Biskamp_03} for plasmas), turbulent fluctuations are known to
possess universality properties. 
In homogeneous turbulence driven at a large scale by an external forcing or
instability, the injected energy at $k_{\mathrm{in}}^{-1}$ cascades
down to smaller scales via local interactions at a constant energy 
transfer rate, and is eventually dissipated at small scales
$k_{\mathrm{d}}^{-1}$ by some dissipation mechanisms. In neutral fluid
turbulence, the viscosity ($\mu$) sets the dissipation scale such that
the dissipation ($\propto \mu k_{\mathrm{d}}^{2}$) balances with the 
energy injection power. In the {\it inertial range} $k_{\mathrm{in}} \ll k \ll
k_{\mathrm{d}}$, the wavenumber spectrum of the kinetic energy follows the
well-known Kolmogorov spectrum $k^{-5/3}$. 

Unlike rather simple neutral fluid turbulence, plasma turbulence is
complex because it consists of multiple species (most simply,
electrons and \revproof{single-species} ions), is anisotropic, and has several linear
wave modes. There are numerous energy flow channels in plasmas, and it
is generally unknown which path the energy takes and what
mechanisms effectively work to dissipate the energy. Possible paths of
cascades and dissipation mechanisms in magnetized plasma turbulence are
\revproof{described comprehensively} in~\citet{SchekochihinCowleyDorland_09}. In
plasmas, the energy carried by Alfv\'en waves and compressive modes
reaches the ion kinetic scale $k\rho_{\mathrm{i}}\sim1$ (if
collisions are rare) where kinetic effects start to play
roles\revproof{:} \revise{phase}
space of cascades is extended into velocity space. Kinetic effects
generally lead to non-Maxwellian distribution functions, i.e. structures
in velocity space are generated, which suffer strong collisional
dissipation as the collision operator provides diffusion in velocity
space. Landau damping along the mean magnetic field and perpendicular
phase mixing \revproof{owing} to finite Larmor radius effects are examples of
dissipation mechanisms in kinetic plasmas~\footnote{We note that these are not an
exhaustive list of the dissipation mechanisms in general. \revproof{As} we only
focus on \revproof{low-frequency} phenomena under the gyrokinetic approximation,
higher-frequency physics, such as the fast \revproof{magnetohydrodynamic
(MHD)} wave and the cyclotron resonance, are neglected.}.

Recently, high-performance kinetic simulations and hybrid fluid-kinetic
modelling covering a wide range of dynamical scales \revproof{have} become
feasible~\citep{GroseljCerriNavarro_17}, and increasing numbers of 
numerical studies have been devoted to the study of dissipation in kinetic
turbulence. Full gyrokinetic simulations were carried out to
study electromagnetic turbulence at the \revproof{sub-ion} Larmor regime aimed at
explaining the energy spectra observed in solar winds. The competition
of the ion entropy cascade resulting in ion heating and kinetic Alfv\'en
wave cascade which carries part of the energy into electron heat was
demonstrated~\citep{HowesTenBargeDorland_11,ToldJenkoTenBarge_15}.
Thermal energy partition in wide parameter space was explored
using a hybrid model for applications to accretion flows~\citep{KawazuraBarnesSchekochihin_18}.
In the absence of the kinetic Alfv\'en waves, the ion entropy cascade
primarily works as the ion dissipation mechanisms via nonlinear phase
mixing in the plane perpendicular to the mean field. The electrostatic
plasma turbulence in two dimensions and the nonlinear phase mixing were studied
theoretically~\citep{PlunkCowleySchekochihin_10} and
numerically~\citep{TatsunoDorlandSchekochihin_09,TatsunoBarnesCowley_10,TatsunoPlunkBarnes_12}.
Compressive fluctuations in Alfv\'enic turbulence were found to have
fluid-like spectra because the phase mixing is strongly
suppressed. The {\it fluidisation} is considered as a reason for the
observed scalings in solar
winds~\citep{SchekochihinParkerHighcock_16,MeyrandKanekarDorland_19},
which is shallower compared with the scaling obtained from a kinetic treatment.

Magnetic reconnection in weakly collisional environments also works as a
process of plasma heating via phase mixing. In addition to the
electron heating due to parallel phase mixing in low-$\beta$
plasmas~\citep{LoureiroSchekochihinzocco_13,NumataLoureiro_14} ($\beta$ is the ratio of
plasma pressure and magnetic pressure), it is shown that the 
nonlinear phase mixing becomes a significant \revproof{ion-heating}
mechanism in high-$\beta$ plasmas during
reconnection~\citep{NumataLoureiro_15a}. In the existence of magnetic
field structures, such as the reconnection magnetic field, the
assumption of local interaction \revproof{no longer holds}. By creating
structures, plasmas may dynamically change the energy flow channel, thus
further \revproof{complicating} the dissipation process.

This work is primarily motivated by the challenge to understand how
plasmas convert the macroscopic energy into heat during turbulent
magnetic reconnection in weakly collisional environments.
To study this problem, we develop a method to impose
external turbulence in gyrokinetic magnetic reconnection simulation
using {\tt AstroGK} code~\citep{NumataHowesTatsuno_10} to extend our
previous studies~\citep{NumataLoureiro_14,NumataLoureiro_15a}.
We restrict our consideration to two dimensions for the sake of
simplicity.

Driving turbulence in kinetic models is not \revproof{straightforward}
as it does not directly evolve a macroscopic flow field. In the
Vlasov\revproof{--}Maxwell equations or gyrokinetic-Maxwell equations, a
distribution function coupled with 
electromagnetic fields via Maxwell's equations is advanced according to
the Vlasov or gyrokinetic equation, and the macroscopic flow is
determined by taking the velocity moment of the distribution function. A
method for driving Alfv\'enic turbulence by injecting an external
current in Amp\`ere's law has already been developed in {\tt
AstroGK} code~\citep{TenBargeHowesDorland_14}.
 However,
it is not suitable for studying the magnetic reconnection problem because
the external current may have a significant direct \revproof{effect} on reconnection
dynamics. Instead, we newly develop a method to directly drive flows in the
gyrokinetic equation by adding a term that only pushes specific moments
of a distribution function. In the gyrokinetic model, excited density
fluctuations that are the \revproof{zeroth-order} moment of the distribution
function couple with the electrostatic potential perturbations through the
quasi-neutrality condition, and yield the so-called $\bm{E}\times\bm{B}$
flows. We note that this forcing method is irrespective of the magnetic
field perturbations in the plane perpendicular to the mean magnetic
field and has close correspondence with forcing in two-dimensional
reduced \revproof{MHD} equations.
\revise{\citet{KawazuraSchekochihinBarnes_20} have also implemented a similar
method for exciting slow mode fluctuations by driving the parallel flow
component of the distribution function in {\tt AstroGK} with the
isothermal electron model. This method also drives the parallel current, and
is not appropriate for the magnetic reconnection study.} 

We also demand the forcing method to have a property that it injects
energy into the system at a constant rate, thus enabling \revproof{the
measurement of} how much of the injected energy goes into specific channels in a
steady state.
A method to control the injection power was proposed for neutral
fluid turbulence~\citep{Alvelius_99}. In the proposed method, the power
input is solely determined by the \revproof{force--force} correlation
(but not by the \revproof{velocity--force} correlation), thus it is
pre-determined. We adopt the same method in two-dimensional gyrokinetics.

In the following sections, we first describe the developed forcing
method in gyrokinetics. We perform a test simulation validating that
the forcing injects energy at a constant rate, and 
a statistically steady state is achieved where the injected power is
balanced by developed dissipation. With this forcing we demonstrate
\revproof{that} the code successfully \revproof{simulates} turbulence
from the \revproof{large-scale fluid to small-scale} kinetic
regimes. The theoretically predicted scaling laws in those regimes are
reproduced. We summarize the paper in \revproof{\S}~\ref{sec:summary}.

\section{External forcing in gyrokinetics}
\label{sec:forcing}

The paper aims to develop a random forcing method in
gyrokinetics, which directly corresponds to that in a fluid model.
By taking a certain limit, the gyrokinetic model (briefly summarized in
\revproof{appendix}~\ref{sec:gkm}) is reduced to the reduced \revproof{MHD} model,
which describes evolutions of the vorticity ($\omega$) and 
magnetic flux ($\psi$) in the direction of the mean magnetic field
$\bm{B}_{0}=B_{0}\bm{z}$. In the reduced MHD, a forcing term can be
written as 
\begin{equation}
 \pdf{\omega}{t} = \dots + a,
\end{equation}
where $a=(\rot \bm{F})\cdot\bm{z}$ with $\bm{F}$ being a force. (We omit all
terms other than the forcing for simplicity. The dots indicate the
omission.) \revproof{In} gyrokinetics, the flow perpendicular to the mean
field is given by the 
$\bm{E}\times\bm{B}$ drift, therefore the vorticity becomes
$\omega=\nabla_{\perp}^{2}\phi/B_{0}$ ($\phi$ is the electrostatic
potential, $\nabla_{\perp}^{2}=\p_{x}^{2}+\p_{y}^{2}$ is the two-dimensional
Laplacian). From the quasi-neutrality condition: 
\begin{equation}
 \sum_{s} \left(\frac{q_{s}^{2}n_{0s}}{T_{0s}} \right) \phi
 = \sum_{s} q_{s} \int h_{s} \diff \bm{v}
 = \sum_{s} q_{s} \delta n_{s},
\end{equation}
we find
\begin{equation}
 \pdf{\omega}{t} = \frac{1}{B_{0}} \nabla_{\perp}^{2} \pdf{\phi}{t}
  = \frac{1}{B_{0}}
  \frac{1}{\sum_{s}\left(q_{s}^{2}n_{0s}/T_{0s}\right)}
  \nabla_{\perp}^{2} \sum_{s} q_{s} \pdf{\delta n_{s}}{t}.
\end{equation}
Note that $\delta n_{s}$ is the density perturbation without the
Boltzmann response part. Therefore, if we add a term in the gyrokinetic
equation that excites density perturbations in an appropriate form, we
get forcing which drives the $\EB$ flows.

We consider an additional term (which we shall call a {\it forcing term}
though it does not literally mean a force) in the gyrokinetic equation
in the following form:
\begin{align}
 \pdf{g_{\bm{k},s}}{t} & = \dots + A_{\bm{k},s},
 \label{eq:forcing_form}
 \\
 A_{\bm{k},s} & = f_{0s} \Xi_{\bm{k},s}(\bm{v})
 \Phi_{\bm{k}}.
 \label{eq:forcing_form_A}
\end{align}
Again, we omit all terms other than the forcing term. \revproof{Here} $f_{0s}$ is a
Maxwellian distribution function of a species $s$\revise{,
$\Phi_{\bm{k}}$ is related to a forcing profile in the $x$--$y$ plane,
$a_{\bm{k}}$, as will be discussed shortly}. For later
convenience, we consider the equation in Fourier space. The velocity
dependence ($\Xi$) is given similar to the Hermite polynomials
\begin{equation}
 \Xi_{\bm{k},s}(\bm{v}) = e^{\frac{b_{s}}{2}}
  \left(
   \frac{N_{s}^{\mathrm{f}}}{n_{0s}} + \frac{T_{s}^{\mathrm{f}}}{T_{0s}}
   \left(
    \frac{v_{\perp}^{2}}{v_{\mathrm{th},s}^{2}} -1 + \frac{b_{s}}{2}
   \right)
  \right),
\end{equation}
where $b_{s}=(k \rho_{s})^{2}/2$, $k=|\bm{k}|$ and $N_{s}^{\mathrm{f}}$,
$T_{s}^{\mathrm{f}}$ are constant coefficients. The gyrokinetic
correction (the terms related to $b_{s}$) is included to compensate for
the $k$ dependence of the velocity integral of $\Xi$.
We intentionally choose no $v_{\parallel}$ dependence of $\Xi$. \revise{(
\citet{KawazuraSchekochihinBarnes_20} \revproof{only included} a term proportional to
$v_{\parallel}$.)} \revproof{Owing} to
this specific choice, the forcing term does not excite $A_{\parallel}$
\revise{and parallel current} perturbations. If $A_{\parallel}=0$ initially, it remains
zero. Therefore, we ignore $A_{\parallel}$ throughout the paper.

By plugging \eqref{eq:forcing_form} into the time derivatives of the
field equations, we obtain the expression for the electromagnetic fields
induced by the forcing term as
\begin{align}
 \revproof{{\mathsfit Y}}_{1} \pdf{\phi_{\bm{k}}}{t} - \revproof{{\mathsfit Y}}_{3}
 \pdf{}{t} \left(\frac{\delta B_{\parallel,\bm{k}}}{B_{0}}\right)
 & = \dots +\Phi_{\bm{k}} X_{N},
 \label{eq:phi_bpar_evol1}\\
 \revproof{{\mathsfit Y}}_{3} \pdf{\phi_{\bm{k}}}{t} + 
 \revproof{{\mathsfit Y}}_{2} \pdf{}{t} \left(\frac{\delta B_{\parallel,\bm{k}}}{B_{0}}\right)
 & = \dots - \Phi_{\bm{k}} X_{P},
 \label{eq:phi_bpar_evol2}
\end{align}
where we have defined the following parameters for convenience,
\begin{align}
 X_{N} & = \sum_{s} q_{s} N_{s}^{\mathrm{f}}, \\
 X_{P} & = \sum_{s} \left(T_{0s} N_{s}^{\mathrm{f}} + n_{0s}T_{s}^{\mathrm{f}}\right),
\end{align}
\revise{and the coefficients of the field equations, $\revproof{{\mathsfit
Y}}_{1,2,3}$, are given in \revproof{appendix}~\ref{sec:gkm}.}
The vorticity equation is derived by solving \eqref{eq:phi_bpar_evol1},
\eqref{eq:phi_bpar_evol2} for $\p \phi_{\bm{k}}/\p t$
\begin{equation}
 \pdf{\omega_{\bm{k}}}{t} = \pdf{}{t} \left(-\frac{k^{2}\phi_{\bm{k}}}{B_{0}}\right)
 = \dots + \frac{1}{B_{0}}
 \left(-k^{2}\Phi_{\bm{k}}\right)
 \frac{\revproof{\mathsfit Y}_{2}X_{N}-\revproof{\mathsfit Y}_{3}X_{P}}
 {\revproof{\mathsfit Y}_{1}\revproof{\mathsfit Y}_{2}+\revproof{\mathsfit Y}_{3}^{2}}.
\end{equation}
Given the desired form of forcing $a_{\bm{k}}$, we have
\begin{equation}
 \Phi_{\bm{k}} = \frac{-a_{\bm{k}}}{k^{2}} B_{0}
 \frac{\revproof{\mathsfit Y}_{1}\revproof{\mathsfit
 Y}_{2}+\revproof{\mathsfit Y}_{3}^{2}}
 {\revproof{\mathsfit Y}_{2}X_{N}-\revproof{\mathsfit Y}_{3}X_{P}}.
\end{equation}

We specify a spatial profile of the forcing by $a_{\bm{k}}$, and a
velocity space profile by $N_{s}^{\mathrm{f}}$ and $T_{s}^{\mathrm{f}}$
determining how density and temperature perturbations are excited.
For $a_{\bm{k}}$, we may borrow the forcing developed for
two-dimensional neutral fluid turbulence exemplified
in~\citet{Carnevale_06}.

\subsection{Injected energy-like quantities}

The generalized energy per unit volume is given by
\begin{equation}
 \overline{W} = \sum_{s} \overline{K}_{s} + \overline{M} =
  \sum_{\bm{k}} \left[
		 \sum_{s} \int \frac{T_{0s} |\delta f_{\bm{k},s}|^{2}}{2f_{0s}}
		 \diff \bm{v}
		 + \frac{|B_{\bm{k}}|^{2}}{2\mu_{0}}
		\right],
\end{equation}
where the bar denotes a quantity per unit volume.
We denote the particle energy by $K_{s}$ and the magnetic energy by $M$.
(We only consider the parallel component of the magnetic energy because
the forcing term does not invoke the perpendicular component.)
By taking the time derivative, we obtain
\begin{align}
 \odf{\overline{K}_{s}}{t} & = \dots + T_{0s}
 \sum_{\bm{k}} \Real \left(\Phi_{\bm{k}}^{\ast}
 \left[
 \int h_{\bm{k},s} \Xi_{\bm{k},s} \diff \bm{v}
 - \frac{1}{T_{0s}} \frac{\revproof{\mathsfit Y}_{3}X_{N}+\revproof{\mathsfit
 Y}_{1}X_{P}}{\revproof{\mathsfit Y}_{1}\revproof{\mathsfit Y}_{2}+\revproof{\mathsfit Y}_{3}^{2}}
 \delta P_{\perp\perp,\bm{k},s}
 \right.\right.
 \nonumber \\
 & ~~~
 - \left.\left.\left[
 N_{s}^{\mathrm{f}}
 - \frac{q_{s}n_{0s}}{T_{0s}}
 \left(1 - \Gamma_{0s}\right)
 \frac{\revproof{\mathsfit Y}_{2}X_{N}-\revproof{\mathsfit Y}_{3}X_{P}}{\revproof{\mathsfit
 Y}_{1}\revproof{\mathsfit Y}_{2}+\revproof{\mathsfit Y}_{3}^{2}}
 - n_{0s} \Gamma_{1s}
 \frac{\revproof{\mathsfit Y}_{3}X_{N}+\revproof{\mathsfit Y}_{1}X_{P}}{\revproof{\mathsfit
 Y}_{1}\revproof{\mathsfit Y}_{2}+\revproof{\mathsfit Y}_{3}^{2}}
 \right]
 \frac{q_{s}\phi_{\bm{k}}}{T_{0s}}
 \right]
 \right),
 \label{eq:part_ene_deriv} \\
 \odf{\overline{M}}{t} & = \dots - 
 \frac{B_{0}^{2}}{\mu_{0}}
 \sum_{\bm{k}}
 \frac{\revproof{\mathsfit Y}_{3}X_{N}+\revproof{\mathsfit Y}_{1}X_{P}}{\revproof{\mathsfit
 Y}_{1}\revproof{\mathsfit Y}_{2}+\revproof{\mathsfit Y}_{3}^{2}}
 \Real \left( \Phi_{\bm{k}}^{\ast}
 \frac{\delta B_{\parallel,\bm{k}}}{B_{0}}\right),
 \label{eq:mag_ene_deriv}
\end{align}
where the asterisk ($^{\ast}$) denotes the complex conjugate and the
pressure perturbation defined by
\begin{equation}
 \delta P_{\perp\perp,\bm{k},s} = \int m_{s} v_{\perp}^{2}
 \frac{J_{1s}}{\alpha_{s}}
 h_{\bm{k},s} \diff \bm{v}
\end{equation}
satisfies the pressure balance relation
\begin{equation}
 \frac{B_{0}^{2}}{\mu_{0}} \frac{\delta B_{\parallel,\bm{k}}}{B_{0}}
 + \sum_{s} \delta P_{\perp\perp,\bm{k},s} = 0.
\end{equation}
Defining the power input $\overline{P}$ by $\diff \overline{W}/\diff t =
\overline{P}$ (ignoring the collisional dissipation), we \revproof{obtain}
\begin{equation}
 \overline{P} = 
 \sum_{\bm{k}} \Real
 \left( \Phi_{\bm{k}}^{\ast}
 \left[
 \sum_{s} T_{0s} \int h_{\bm{k},s} \Xi_{\bm{k},s} \diff \bm{v}
 \right]
 \right).
\end{equation}
The coefficient of the $q_{s}\phi_{\bm{k}}/T_{0s}$ term in
\eqref{eq:part_ene_deriv} vanishes when the species sum is taken.

In the electrostatic limit, there is an additional invariant, which we
call the electrostatic invariant $E$,
\begin{equation}
 \overline{E} = \frac{1}{2} \sum_{\bm{k}}
  \left[ 
   \sum_{s} \frac{q_{s}^{2}n_{0s}}{T_{0s}} \left(1 - \Gamma_{0s} \right)
  \right]
  \left| \phi_{\bm{k}}\right|^{2}.
\end{equation}
The change of the electrostatic invariant due to the forcing is given by
\begin{equation}
 \odf{\overline{E}}{t} = \dots
  + X_{N} \sum_{\bm{k}} \Real
  \left(\Phi_{\bm{k}}^{\ast}\phi_{\bm{k}}\right).
\end{equation}

\subsection{Numerical implementation of constant power injection}

  We simply add the forcing term in the {\tt AstroGK} code as
  \begin{equation}
   g^{n+1}_{\bm{k},s} = g^{n}_{\bm{k},s} + \Delta t A_{\bm{k},s},
  \end{equation}
  where the superscript denotes the time step.
  It leads to an increment of $\Delta h_{\bm{k},s} = h_{\bm{k},s}^{n+1}
  - h_{\bm{k},s}^{n}$ as
  \begin{equation}
   \Delta h_{\bm{k},s}
   =
   \Delta t A_{\bm{k},s} 
   + \frac{q_{s} \Delta \phi_{\bm{k}}}{T_{0s}} J_{0s} f_{0s}
   + \frac{2v_{\perp}^{2}}{v_{\mathrm{th},s}^{2}}
   \frac{J_{1s}}{\alpha_{s}} f_{0s}
   \frac{\Delta \delta B_{\parallel,\bm{k}}}{B_{0}}.
  \end{equation}
  The increments of the fields are obtained from the discretized field
  equations \eqref{eq:phi_bpar_evol1} \revproof{and} \eqref{eq:phi_bpar_evol2},
  \begin{align}
   \begin{pmatrix}
    \Delta \phi_{\bm{k}} \\ 
    \Delta \delta B_{\parallel,\bm{k}} / B_{0}
   \end{pmatrix}
   & = 
   \begin{pmatrix}
    \phi_{\bm{k}}^{n+1} - \phi_{\bm{k}}^{n} \\
    \delta B_{\parallel,\bm{k}}^{n+1} / B_{0} - \delta B_{\parallel,\bm{k}}^{n} / B_{0}
   \end{pmatrix}
   \nonumber \\
   & = \frac{1}{\revproof{\mathsfit Y}_{1}\revproof{\mathsfit Y}_{2}+\revproof{\mathsfit Y}_{3}^{2}}
   \begin{pmatrix}
    \revproof{\mathsfit Y}_{2} & \revproof{\mathsfit Y}_{3}\\
    - \revproof{\mathsfit Y}_{3} & \revproof{\mathsfit Y}_{1}
   \end{pmatrix}
   \begin{pmatrix}
    X_{N} \\ - X_{P}
   \end{pmatrix}
   \Phi_{\bm{k}} \Delta t.
  \end{align}

  The power input $\overline{P}=(\overline{W}^{n+1}-\overline{W}^{n})/\Delta t$ becomes
  \begin{equation}
   \overline{P} = 
    \sum_{\bm{k}} \Real
    \left( \Phi_{\bm{k}}^{\ast}
     \left[
      \sum_{s} T_{0s} \int h_{\bm{k},s}^{n+1/2} \Xi_{\bm{k},s} \diff \bm{v}
     \right]
    \right),
 \label{eq:power}
  \end{equation}
  where the superscript $n+1/2$ means the average of the values at the time
  steps $n$ and $n+1$. This is decomposed into two components
  $\overline{P}=\overline{P}_{1}+\Delta t \overline{P}_{2}$, and
  \begin{align}
   \overline{P}_{1} & =
   \sum_{\bm{k}} \Real
   \left( \Phi_{\bm{k}}^{\ast}
   \left[
   \sum_{s} T_{0s} \int h_{\bm{k},s}^{n} \Xi_{\bm{k},s} \diff \bm{v}
   \right]
   \right) \equiv
   \sum_{\bm{k}} \Real \left(\Phi_{\bm{k}}^{\ast}
   \Psi_{\bm{k}}^{n}\right),
   \label{eq:power1}
   \\
   \overline{P}_{2} & =
   \sum_{\bm{k}}
   \left( 
   \frac{1}{2} \frac{\revproof{\mathsfit Y}_{2}X_{N}^{2} - 2 \revproof{\mathsfit Y}_{3}X_{N}X_{P} -
   \revproof{\mathsfit Y}_{1}X_{P}^{2}}
   {\revproof{\mathsfit Y}_{1}\revproof{\mathsfit Y}_{2}+\revproof{\mathsfit Y}_{3}^{2}}
   \right.
   \nonumber
   \\
   & ~~~
   + \left.
   \sum_{s} \frac{n_{0s}T_{0s}}{2}
   e^{b_{s}}
   \left[
   \left(\frac{N_{s}^{f}}{n_{0s}}\right)^{2}
   + b_{s}
   \left(\frac{N_{s}^{f}}{n_{0s}}\right)
   \left(\frac{T_{s}^{f}}{T_{0s}}\right)
   + \left(1+\frac{b_{s}^{2}}{4}\right)
   \left(\frac{T_{s}^{f}}{T_{0s}}\right)^{2}
   \right]
   \right)
   \left|\Phi_{\bm{k}}\right|^{2}
   \nonumber
   \\
   & \equiv
   \sum_{\bm{k}} \Upsilon(k) \left|\Phi_{\bm{k}}\right|^{2}.
   \label{eq:power2}
  \end{align}
In the electrostatic case, $\delta B_{\parallel}=0$, and
\begin{equation}
 \Upsilon(k) = \frac{1}{2}
 \left( \frac{X_{N}^{2}}{\revproof{\mathsfit Y}_{1}} + 
   \sum_{s} n_{0s}T_{0s}
   e^{b_{s}}
   \left[
   \left(\frac{N_{s}^{f}}{n_{0s}}\right)^{2}
   + b_{s}
   \left(\frac{N_{s}^{f}}{n_{0s}}\right)
   \left(\frac{T_{s}^{f}}{T_{0s}}\right)
   + \left(1+\frac{b_{s}^{2}}{4}\right)
   \left(\frac{T_{s}^{f}}{T_{0s}}\right)^{2}
   \right]
 \right).
\end{equation}

In general, generated fields by random forcing cannot be controlled,
therefore, the input power is unknown. However, if we can vanish
$\overline{P}_{1}$ by appropriately choosing $\Phi_{\bm{k}}$, the input
power $\overline{P}_{2}$ can be controlled because it is solely
determined by the forcing parameters. Such a method was suggested
by~\citet{Alvelius_99}. Following~\citet{Alvelius_99}, we assume the
forcing is isotropic and it only depends on $k$. If we can make
$\overline{P}_{1}=0$, the total input power becomes
\begin{equation}
 \overline{P} = \Delta t \overline{P}_{2} = 
 \Delta t \sum_{\bm{k}} \Upsilon(k) |\Phi_{\bm{k}}|^{2}.
 \label{eq:p_wo_p1}
\end{equation}
To express the isotropy, we write $\bm{k}$ in the polar coordinate
$(k,\theta_{k})$, and introduce the shell average in $\theta_{k}$ as
follows
\begin{align}
 \left\langle \Upsilon(k) \left|\Phi_{\bm{k}}\right|^{2} \right\rangle_{\theta_{k}}
 & = \frac{1}{N_{k}} \sum_{k\leq |\bm{k}'| \leq k+\Delta k} \Upsilon(k')
 \left|\Phi_{\bm{k}'}\right|^{2}, \\
 \sum_{\bm{k}} \Upsilon(k) \left|\Phi_{\bm{k}}\right|^{2}
 & = \sum_{|\bm{k}|} N_{k} \left\langle \Upsilon(k) \left|\Phi_{\bm{k}}\right|^{2} \right\rangle_{\theta_{k}},
\end{align}
where $N_{k}$ \revproof{denotes} the number of discrete Fourier
modes in the $k$ shell. \revproof{As} $N_{k}$ is proportional to $2\pi
k$, we write $N_{k}=c_{1} k$ with $c_{1}$ being a constant. If
$\Phi_{\bm{k}}$ does not depend on $\theta_{k}$, we can omit the shell
average. Then, the power input is given by
\begin{equation}
 \overline{P}
 = \Delta t \sum_{|\bm{k}|} c_{1} k \Upsilon(k)
 \left|\Phi_{\bm{k}}\right|^{2} 
 = \Delta t \sum_{|\bm{k}|} c_{1} \Upsilon(k) \frac{B_{0}^{2}\left(\revproof{\mathsfit Y}_{1}\revproof{\mathsfit Y}_{2} + \revproof{\mathsfit Y}_{3}^{2}\right)^2}{\left(\revproof{\mathsfit Y}_{2} X_{N} -
							      \revproof{\mathsfit Y}_{3} X_{P}\right)^{2}} 
 \frac{\left|a_{\bm{k}}\right|^{2}}{k^{3}}.
\end{equation}

For a spectral profile of $a_{\bm{k}}$, we want the energy
isotropically injected only on a large scale. Therefore, we assume a
forcing having \revproof{the} following two-dimensional isotropic
Gaussian profile in the wavenumber space,
\begin{equation}
 \overline{P} = c_{2} \sum_{|\bm{k}|}
  \revproof{\exp}\left(-\left(\frac{k-k_{\mathrm{in}}}{k_{\mathrm{w}}}\right)^{2}\right),
\end{equation}
where $c_{2}$, $k_{\mathrm{in}}$ and $k_{\mathrm{w}}$ are
constants. Then, the amplitude of $a_{\bm{k}}$ becomes
\begin{equation}
 \left| a_{\bm{k}} \right| =
  \left(\frac{1}{\Delta t}
   \frac{\left(\revproof{\mathsfit Y}_{2}X_{N}-\revproof{\mathsfit
	  Y}_{3}X_{P}\right)^{2}}{B_{0}^{2}(\revproof{\mathsfit Y}_{1}\revproof{\mathsfit
   Y}_{2}+\revproof{\mathsfit Y}_{3}^{2})^{2}}
   \frac{k^{3}}{\Upsilon(k)}
 \frac{c_{2}}{c_{1}}
 \revproof{\exp}\left(-\left(\frac{k-k_{\mathrm{in}}}{k_{\mathrm{w}}}\right)^{2}\right)
 \right)^{1/2}.
\end{equation}
By plugging this expression into \eqref{eq:p_wo_p1} and denoting the
input power by $\overline{P}_{\mathrm{in}}$, we determine the unknown
constant coefficients as
\begin{equation}
 \frac{c_{2}}{c_{1}} = \frac{\overline{P}_{\mathrm{in}}}{\sum_{\bm{k}} \revproof{\exp}\left(-\left(\frac{k-k_{\mathrm{in}}}{k_{\mathrm{w}}}\right)^{2}\right)/k}.
\end{equation}

The method proposed in~\citet{Alvelius_99} is that the phase of forcing
is chosen such that $P_{1}$ vanishes. We write
\begin{equation}
 a_{\bm{k}} = \left|a_{\bm{k}}\right| e^{\imag(\revise{\varsigma_{1}}+\revise{\varsigma_{2}})},
  \label{eq:ak_prof}
\end{equation}
where $\revise{\varsigma_{2}}=2\upi X$ with $X\in[0,1)$ being a uniformly
distributed random number, and \revise{$\varsigma_{1}$} is chosen to
vanish $P_{1}$. To diminish $\Real\left(\Phi_{\bm{k}}^{\ast} \Psi_{\bm{k}}\right)$, we
demand
\begin{equation}
 \tan \revise{\varsigma_{1}} = \frac{
  \Real\left(\Psi_{\bm{k}}\right)\cos\revise{\varsigma_{2}} +
  \Imag\left(\Psi_{\bm{k}}\right)\sin\revise{\varsigma_{2}}
  }
  {
  \Real\left(\Psi_{\bm{k}}\right)\sin\revise{\varsigma_{2}} - 
  \Imag\left(\Psi_{\bm{k}}\right)\cos\revise{\varsigma_{2}}
  }.
\end{equation}

In summary, the form of forcing is given as follows
\begin{align}
 a_{\bm{k}}
 & = 
  \left(\frac{\overline{P}_{\mathrm{in}}}{\Delta t}
   \frac{\left(\revproof{\mathsfit Y}_{2}X_{N}-\revproof{\mathsfit
 Y}_{3}X_{P}\right)^{2}}{B_{0}^{2}(\revproof{\mathsfit Y}_{1}\revproof{\mathsfit
 Y}_{2}+\revproof{\mathsfit Y}_{3}^{2})^{2}}
   \frac{k^{3}}{\Upsilon(k)}
 \frac{\revproof{\exp}\left(-\left(\frac{k-k_{\mathrm{in}}}{k_{\mathrm{w}}}\right)^{2}\right)}
 {\sum_{\bm{k}} \revproof{\exp}\left(-\left(\frac{k-k_{\mathrm{in}}}{k_{\mathrm{w}}}\right)^{2}\right)/k}
 \right)^{1/2}
 e^{\imag \left(\revise{\varsigma_{1}}+\revise{\varsigma_{2}}\right)},
 \\
 \revise{\varsigma_{1}} & = \tan^{-1} \left(
 \frac{
 \Real\left(\Psi_{\bm{k}}\right)\cos\revise{\varsigma_{2}} +
 \Imag\left(\Psi_{\bm{k}}\right)\sin\revise{\varsigma_{2}}
 }
 {
 \Real\left(\Psi_{\bm{k}}\right)\sin\revise{\varsigma_{2}} - 
 \Imag\left(\Psi_{\bm{k}}\right)\cos\revise{\varsigma_{2}}
 }
 \right).
\end{align}
One generates a random phase \revise{$\varsigma_{2}$} at every time step, and
calculates \revise{$\varsigma_{1}$} from $\Psi_{\bm{k}}$ which depends on the
instantaneous value of $h_{\bm{k}}$.
Arbitrary input parameters are the input power $\overline{P}_{\mathrm{in}}$,
the injection scale $k_{\mathrm{in}}$, the injection range
$k_{\mathrm{w}}$ and $N_{s}^{\mathrm{f}}$, $T_{s}^{\mathrm{f}}$
controlling the velocity space profile.
\revise{We note that it is straightforward to apply the same method to
control the injection of electrostatic invariant. However, either the
injection of $W$ or $E$ can be fixed, but not both at the same time.}

\subsection{Validation}
\label{sec:validation}

We have implemented the forcing term in {\tt AstroGK}. In this section,
we test the term has the intended properties.
We denote the box size $L_{x}=L_{y}=2\upi L$, and set the ion Larmor
radius $\rho_{\mathrm{i}}/L=0.01$. Both ions and electrons are kinetic.
The ratios of mass, charge, background density and background temperature are
$m_{\mathrm{i}}/m_{\mathrm{e}}=100$, $q_{\mathrm{i}}/q_{\mathrm{e}}=-1$,
$n_{0\mathrm{i}}/n_{0\mathrm{e}}=1$ \revproof{and}
$T_{0\mathrm{i}}/T_{0\mathrm{e}}=1$.
We set $\beta_{\mathrm{i,e}}=1$, and include magnetic fluctuations.
The parameters for the forcing are
\begin{align}
 k_{\mathrm{in}}L & = 2, &
 k_{\mathrm{w}} L & = 1, \\
 N_{\mathrm{i}}^{\mathrm{f}}/n_{0\mathrm{i}} & = 1, &
 T_{\mathrm{i}}^{\mathrm{f}}/T_{0\mathrm{i}} & = 1.
\end{align}
The power input defines the time scale of the system. If we assume that the
injected energy is deposited to the ion kinetic energy
$\overline{K}_{\mathrm{in}}=n_{0\mathrm{i}} m_{\mathrm{i}} u_{\mathrm{in}}^{2}/2 =
\overline{P}_{\mathrm{in}} \tau_{\mathrm{in}}$, and
define the characteristic time at the input scale as
$\tau_{\mathrm{in}}=1/(k_{\mathrm{in}}u_{\mathrm{in}})$, we obtain
 \begin{equation}
  \tau_{\mathrm{in}} = \left(\frac{n_{0\mathrm{i}}m_{\mathrm{i}}}
			{2\overline{P}_{\mathrm{in}} k_{\mathrm{in}}^{2}}\right)^{1/3}.
 \end{equation}

\begin{figure}
 \begin{center}
  \includegraphics[scale=0.8]{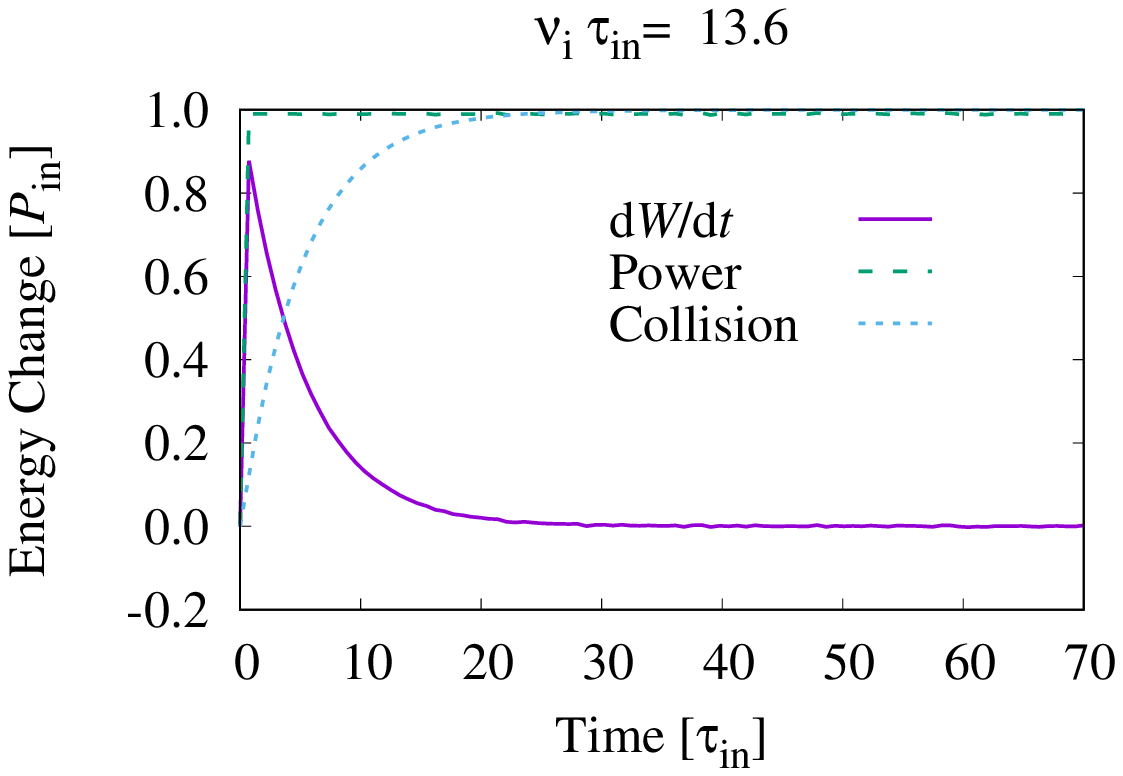}
  \includegraphics[scale=0.8]{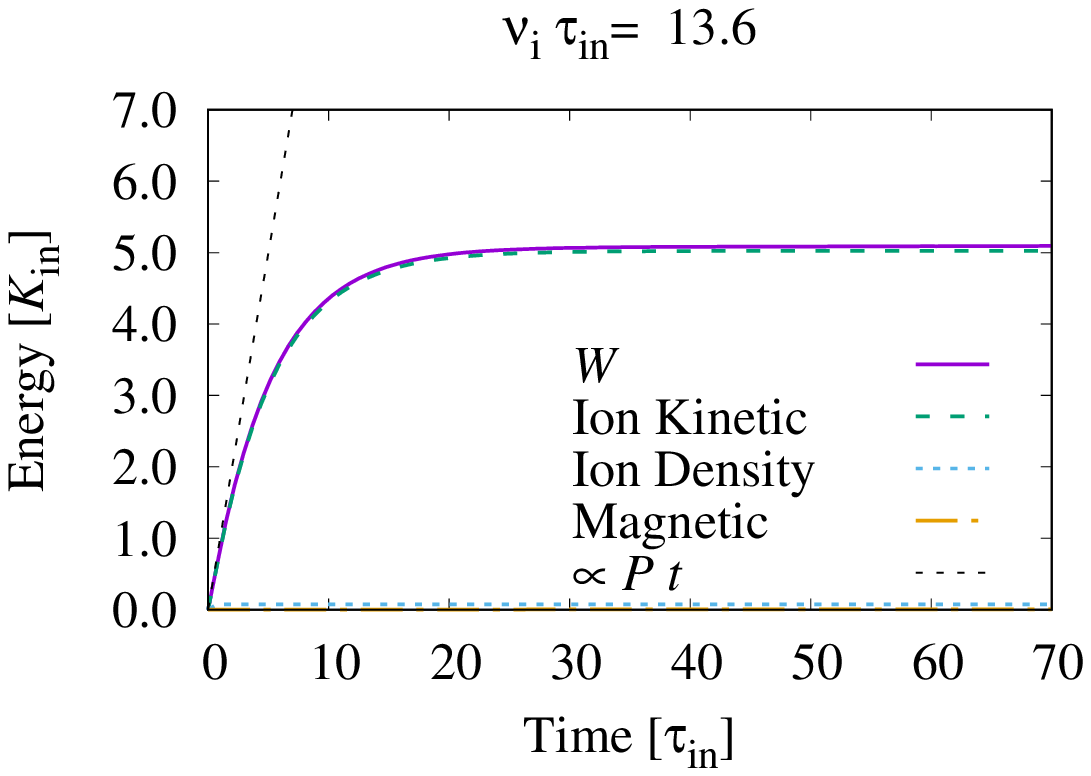}
  \caption{\label{fig:powbal}Power balance and energy evolution. The
  system reaches a statistically steady state where the energy input and
  collisional dissipation are balanced.}
 \end{center}
\end{figure}

Figure~\ref{fig:powbal} shows the power balance (top) and total energy
evolution (bottom). The injected power $\overline{P}$ measured by the formula
\eqref{eq:power} stays at the given constant value $\overline{P}_{\mathrm{in}}$
throughout the run, which confirms the method described in
\revproof{\S}~\ref{sec:forcing} is correctly implemented and works as
desired. \revproof{Owing to} the finite collisionality, the collisional
energy dissipation initially increases and reaches the same level as the
input, where a
statistically steady state is achieved. The energy is mostly dominated
by the ion kinetic energy, $\sum_{\bm{k}} m_{\mathrm{i}} n_{0\mathrm{i}}
|\bm{u}_{\perp,\mathrm{i},\bm{k}}|^{2}/2$, \revproof{whereas} the density variance of ions,
$\sum_{\bm{k}} n_{0\mathrm{i}}T_{0\mathrm{i}}
\left|n_{\mathrm{i},\bm{k}}/n_{0\mathrm{i}}\right|^2/2$ (where
$n_{\mathrm{i},\bm{k}}/n_{0\mathrm{i}}=\delta
n_{\mathrm{i},\bm{k}}/n_{0\mathrm{i}}-q_{\mathrm{i}} \phi_{\bm{k}}/T_{0\mathrm{i}}$), and
the magnetic energy, $\sum_{\bm{k}}\left|\delta B_{\parallel,\bm{k}}^2/(2\mu_{0})\right|$, are
small. The total energy initially increases with the energy input rate
$\overline{P}_{\mathrm{in}}$,
and saturates at around 20 eddy \revproof{times}. Note that we set the collision
frequency $\nu_{\mathrm{ii,ee}}\tau_{\mathrm{in}}\approx13.6$, therefore
particles have experienced sufficient collisions before reaching the
steady state. The time to reach the steady state and the saturated
energy level depend on how \revproof{the} turbulent spectrum develops,
thus are not known {\it a priori}. In the current setup, the
collisional dissipation is mostly provided by
electrons. \revproof{Thus}, if electrons are not kinetic, dissipation
does not develop: \revproof{the} total energy of 
the system unlimitedly increases and a steady state is not achieved.
The behaviour of turbulent spectra and how
dissipation develops are of central importance to the turbulence study,
which we discuss in the next section.

\section{Scaling laws of driven gyrokinetic turbulence\revproof{:} from fluid to
  kinetic regimes}

In this section, we demonstrate scaling laws of turbulent fluctuations
in both fluid ($k\rho_{\mathrm{i}}\ll1$) and kinetic
($k\rho_{\mathrm{i}}\sim1$) regimes. The theory of gyrokinetic turbulent
cascade was \revproof{described comprehensively}
in~\citet{SchekochihinCowleyDorland_09}. We 
only consider the case without (kinetic) Alfv\'en waves and in two
dimensions. \citet{PlunkCowleySchekochihin_10} specifically focused on gyrokinetic
turbulence in two dimensions, and discussed the relation with the
Charney\revproof{--}Hasegawa\revproof{--}Mima \revproof{(CHM)} model in the long-wavelength
fluid regime. After briefly reviewing the scaling laws derived in those
works, we show numerical simulations of driven gyrokinetic turbulence
and test if those scalings are established in statistically steady
states. We note some essential aspects of the electrostatic gyrokinetic
turbulence were already shown in decaying turbulence
simulations~\citep{TatsunoDorlandSchekochihin_09,TatsunoBarnesCowley_10,TatsunoPlunkBarnes_12}.

In the system considered in this paper, we ignore magnetic perturbation
perpendicular to the mean field (no kinetic Alfv\'en waves). Therefore,
the macroscopic system is described by the \revproof{CHM}
equation (or Navier-Stokes equation in some limit). In
two-dimensional turbulence, a dual cascade of the energy and enstrophy
leads to  the energy spectrum of $k^{-3}$ in the inertial
range. For a high-$\beta$ case, also compressive fluctuations exist
which are passively advected by the flow. A transition occurs at the ion kinetic
scale. In the kinetic regime, the so-called nonlinear phase mixing
starts \revproof{to play a role}, and the behaviour of turbulence changes where turbulent
cascades also occur in velocity space. The energy further cascades down
to even smaller scales where collisions eventually take out the energy
from the system. How the energy is dissipated depends on the particular
models of electrons.

\subsection{Inertial range}

In the \revproof{long-wavelength} limit $k\rho_{\mathrm{i}}\ll1$, the gyro-average
acts as unity and
$\Gamma_{0}(k^{2}\rho_{\mathrm{i}}^{2}/2)\approx1-k^{2}\rho_{\mathrm{i}}^{2}/2$. If
we further assume that $\beta$ is low (the electrostatic limit) and ions
are cold, we reach the well-known \revproof{CHM}
equation~\citep{Charney_71,HasegawaMima_77},
\begin{equation}
 \pdf{}{t} \left(2Q+k^{2}\rho_{\mathrm{i}}^{2} \right) \phi_{\bm{k}} - \frac{1}{B_{0}}
  {\mathcal F}
  \left(\left\{\phi,\rho_{\mathrm{i}}^{2}\nabla^{2}\phi\right\}\right)
  = {\mathcal C}_{\bm{k}} + {\mathcal A}_{\bm{k}},
 \label{eq:chm}
\end{equation}
where ${\mathcal C}_{\bm{k}} =
2T_{0\mathrm{i}}/(q_{\mathrm{i}}n_{0\mathrm{i}}) \int C_{\bm{k}} \diff
\bm{v}$, ${\mathcal A}_{\bm{k}} =
2T_{0_{\mathrm{i}}}/(q_{\mathrm{i}}n_{0\mathrm{i}}) \int A_{\bm{k}}
\diff\bm{v}$. The parameter $Q=Q_{\mathrm{e}}/Q_{\mathrm{i}}$
($Q_{s}=q_{s}^{2}n_{0\mathrm{s}}/T_{0s}$) determines the electron
response. The cold ion assumption demands $Q \sim k^{2} \rho_{\mathrm{i}}^{2}$. 
We immediately find that there are two invariants, the energy and
enstrophy, by multiplying \eqref{eq:chm} by
$\phi_{\bm{k}}^{\ast}/\rho_{\mathrm{i}}^{2}$ and
$k^{2}\phi_{\bm{k}}^{\ast}/\rho_{\mathrm{i}}^{2}$,
\begin{align}
 \overline{E}_{\mathrm{CHM}} & = \frac{n_{0\mathrm{i}}m_{\mathrm{i}}}{2B_{0}^{2}} \sum_{\bm{k}}
 \left(\frac{2Q}{\rho_{\mathrm{i}}^{2}}\left|\phi_{\bm{k}}\right|^{2}
 + k^{2}\left|\phi_{\bm{k}}\right|^{2}\right),
 \\
 \overline{Z}_{\mathrm{CHM}} & =
 \frac{n_{0\mathrm{i}}m_{\mathrm{i}}}{2B_{0}^{2}} \sum_{\bm{k}}
 \left( \frac{2Q}{\rho_{\mathrm{i}}^{2}} k^{2} \left|\phi_{\bm{k}}\right|^{2} +
 k^{4} \left|\phi_{\bm{k}}\right|^{2}\right).
\end{align}
\revproof{Owing} to the conservation of energy and enstrophy, a dual cascade will
occur in two dimensions: the energy inversely cascades to a larger scale
$k<k_{\mathrm{in}}$, \revproof{whereas} the enstrophy cascades to a smaller scale
$k>k_{\mathrm{in}}$. 

By assuming isotropy and locality of nonlinear interactions, the CHM
equation leads to the following relation at each scale
$\ell$,
\begin{equation}
 \frac{1}{\tau_{\ell}} \left( 2Q + \frac{\rho_{\mathrm{i}}^{2}}{\ell^{2}} \right)
  \sim \frac{1}{B_{0}} \frac{\rho_{\mathrm{i}}^{2} \phi_{\ell}}{\ell^{4}},
  \label{eq:nonlin_scaling_phi_fluid}
\end{equation}
where $\tau_\ell$ is the nonlinear decorrelation time.
We demand the constancy of the enstrophy flux for the forward cascade:
\begin{equation}
 \frac{1}{\tau_{\ell}} \frac{1}{2B_{0}^{2}}
  \left( \frac{2Q}{\rho_{\mathrm{i}}^{2}}
   \frac{\phi_{\ell}^{2}}{\ell^{2}} + 
   \frac{\phi_{\ell}^{2}}{\ell^{4}}\right)
  \sim \frac{\varepsilon_{Z}}{n_{0\mathrm{i}}m_{\mathrm{i}}} = \revproof{\textrm{const}.}
\end{equation}
Combining the two \revproof{relations} yields the scaling for $\phi_{\ell}$
\begin{equation}
 \frac{\phi_{\ell}}{B_{0}} \sim
  \left(\frac{\varepsilon_{Z}}{n_{0\mathrm{i}}m_{\mathrm{i}}}\right)^{1/3} \ell^{2},
\end{equation}
which corresponds to the scaling for the energy and enstrophy
\begin{align}
 \frac{\overline{E}_{\mathrm{CHM},k}}{n_{0\mathrm{i}}m_{\mathrm{i}}}
 & \sim
 \left(\frac{\varepsilon_{Z}}{n_{0\mathrm{i}}m_{\mathrm{i}}}\right)^{2/3}
 \left( \frac{2Q}{\rho_{\mathrm{i}}^{2}} + k^{2} \right) k^{-5},
 \\
 \frac{\overline{Z}_{\mathrm{CHM},k}}{n_{0\mathrm{i}}m_{\mathrm{i}}}
 & \sim
 \left(\frac{\varepsilon_{Z}}{n_{0\mathrm{i}}m_{\mathrm{i}}}\right)^{2/3}
 \left( \frac{2Q}{\rho_{\mathrm{i}}^{2}} + k^{2} \right)k^{-3}.
\end{align}
These scalings indicate that the energy and enstrophy scalings break
around $k\rho_{\mathrm{i}}\sim \sqrt{2Q}$, where the ion polarisation effect
sets in.

In the inverse cascade $k<k_{\mathrm{in}}$, we demand the constancy of
the energy flux, \revproof{$\varepsilon_{E}$},
\begin{equation}
 \frac{1}{\tau_{\ell}} \frac{1}{2B_{0}^{2}}
 \left( \frac{2Q}{\rho_{\mathrm{i}}^{2}} \phi_{\ell}^{2} +
 \frac{\phi_{\ell}^{2}}{\ell^{2}}
 \right)
 \sim \frac{\revproof{\varepsilon_{E}}}{n_{0\mathrm{i}}m_{\mathrm{i}}} = \revproof{\textrm{const.}},
\end{equation}
leading to the scaling
\begin{equation}
 \frac{\phi_{\ell}}{B_{0}} \sim 
 \left( \frac{\revproof{\varepsilon_{E}}}{n_{0\mathrm{i}}m_{\mathrm{i}}}\right)^{1/3}
 \ell^{4/3}.
\end{equation}

\begin{figure}
 \centering
 \includegraphics[scale=0.8]{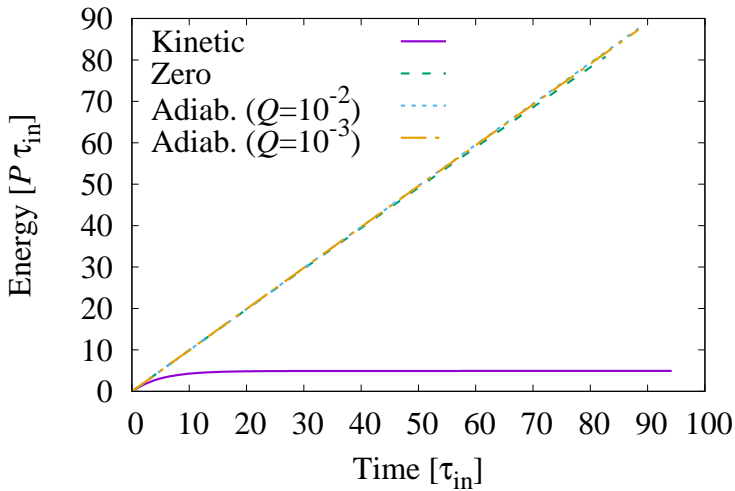}
 \caption{\label{fig:fluid_tevo}Time evolution of the energy for the
 $k_{\mathrm{in}}\rho_{\mathrm{i}}=0.02\ll 1$ case. The electron models are
 gyrokinetic, zero response ($Q=0$) and adiabatic response with
 $Q=10^{-2},10^{-3}$. The total energy saturates only for the
 gyrokinetic electron case.}
\end{figure}

We perform gyrokinetic simulations at $k\rho_{\mathrm{i}}\ll 1$ with the parameters
the same as \revproof{in \S}~\ref{sec:validation} except that here
$\beta_{\mathrm{i}}=0$. Three types of electron response are examined,
namely kinetic, adiabatic ($Q=10^{-2}, 10^{-3}$) and zero response
($Q=0$). Figure~\ref{fig:fluid_tevo} shows the time evolution of the
generalized energy $\overline{W}$. The energy saturates only for
the gyrokinetic electron case, \revproof{whereas} it continuously 
increases during the simulations for other cases.

\begin{figure}
 \centering
 \includegraphics[scale=0.8]{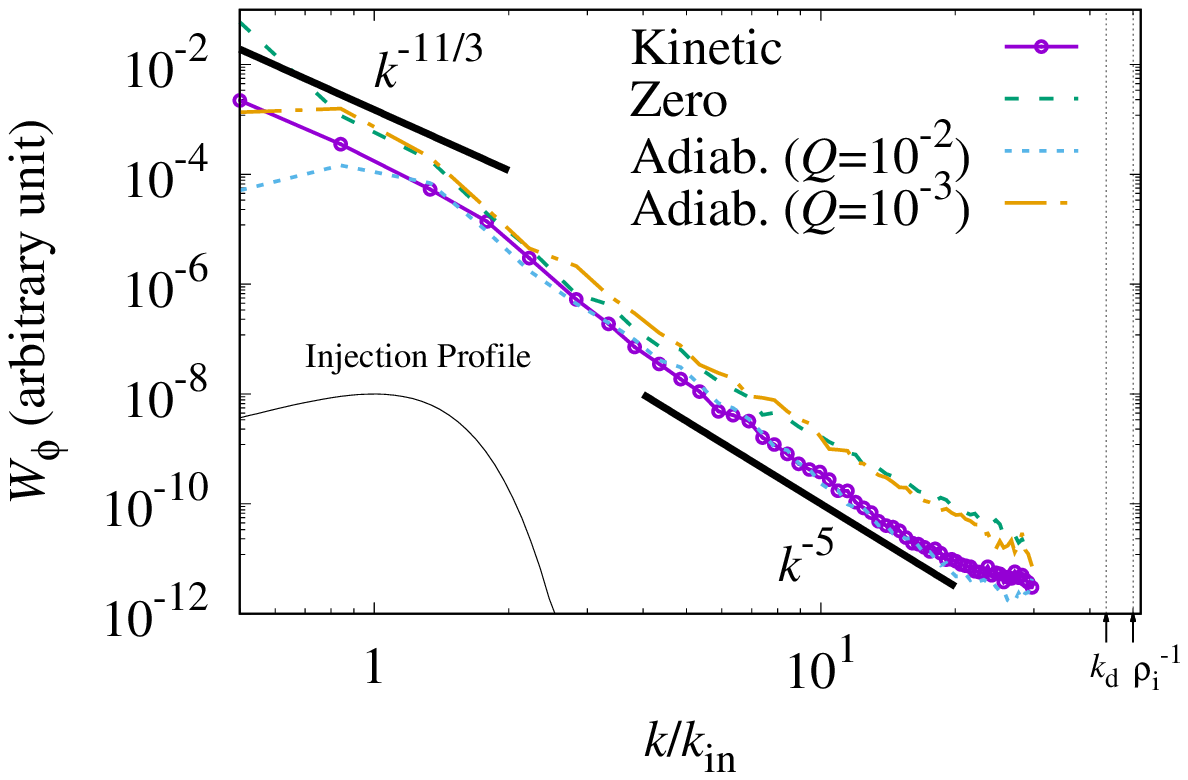}
 \includegraphics[scale=0.8]{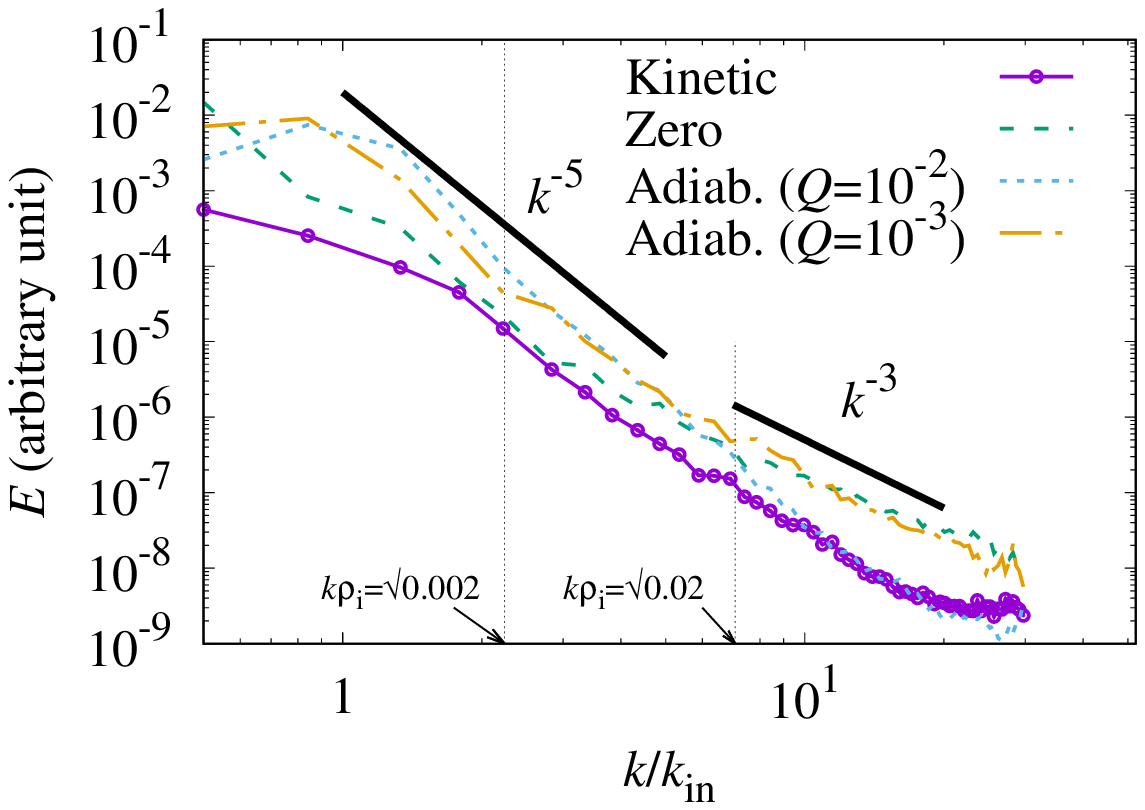}
 \caption{\label{fig:fluid_spec}Spectra of $\overline{W}_{\phi}$ and
 $\overline{E}_{\mathrm{CHM}}$ for the $k_{\mathrm{in}}\rho_{\mathrm{i}}=0.02\ll1$
 case. The gyrokinetic simulation successfully reproduces the scaling
 laws in the fluid regime.}
\end{figure}

We then examine the wavenumber spectra after turbulence is fully developed.
It is convenient to define the spectrum of $\overline{W}_{\phi,k}$
\begin{equation}
 \overline{W}_{\phi,k} \diff k = \frac{q_{\mathrm{i}}^{2}
  n_{0\mathrm{i}}}{2T_{0\mathrm{i}}} \left|\phi_{\bm{k}}\right|^{2}
\end{equation}
to compare the simulations with the theoretical predictions in this regime as it is
irrespective of the electron response. In \revproof{figure}~\ref{fig:fluid_spec}, the
wavenumber spectra of $\overline{W}_{\phi,k}$ and $\overline{E}_{\mathrm{CHM},k}$ are
shown. The spectra are averages of 100 time snapshots to
smooth out temporal variations. The wavenumber is normalized by the
injection scale $k_{\mathrm{in}}$. From the results of $\overline{W}_{\phi,k}$, we
confirm that the theoretical scalings of both the 
forward ($k/k_{\mathrm{in}}>1$) and inverse ($k/k_{\mathrm{in}}<1$)
cascades are successfully reproduced\footnote{\revise{The
injection scale is not well separated from the inverse cascade range in
the current setup. However, it is confirmed, by a separate simulation
run with $k_{\mathrm{in}} L \gg 1$, that the inverse cascade spectrum is
clearly observed}.}.
In the figure showing $\overline{E}_{\mathrm{CHM},k}$, we depict the
transition point $k\rho_{\mathrm{i}}=\sqrt{2Q}$ and the predicted slopes
for $k\rho_{\mathrm{i}}\gtrless \sqrt{2Q}$ for reference. Although the
transition is not sharp, we observe the slopes at higher and lower
wavenumber ranges are different and follow the predictions.

In the non-gyrokinetic electron cases, the energy inversely cascades to
accumulate at a low-$k$ regime. Therefore, there exists no saturation
mechanism. However, the accumulated energy is somehow extracted from the
system if electrons are gyrokinetic. In fact, the main component of
dissipation which balances with the injection is that 
of electrons. Because the kinetic effect is weak in this \revproof{small-}$k$
regime, the electron distribution function is almost Maxwellian, and the
collision operator employed in {\tt
AstroGK}~\citep{AbelBarnesCowley_08,BarnesAbelDorland_09} is dominated by  
the gyro-diffusion term,
\begin{equation}
 C(h_{s}) = - \frac{k^{2} \rho_{s}^{2}}{4}
  \left( \nu_{\mathrm{D}}(1+\xi^{2}) + \nu_{\parallel}
   (1-\xi^{2})\right)
  \frac{v^{2}}{v_{\mathrm{th},s}^{2}} h_s,
\end{equation}
 which acts like friction. Suppose $h_{s}$ is a Maxwellian,
\begin{equation}
 h_{s} = \frac{\delta n_{s}}{n_{0s}} f_{0s},
\end{equation}
the collision term yields
\begin{equation}
 \int C(h_{s}) \diff \bm{v} = - 
  \frac{\sqrt{2}}{3\sqrt{\upi}} \nu_{s} k^{2} \rho_{s}^{2} \delta n_{s},
\end{equation}
which corresponds to a friction force in the vorticity equation.
Therefore, the large scale flow can be decelerated to reach the steady
state if the electron collision is incorporated.

\subsection{Ion entropy cascade}

On a smaller scale, ions follow the gyrokinetic equation where the
nonlinear phase mixing leads to the cascade of entropy. We consider a
scale $\rho_{\mathrm{e}}\ll\ell\ll\rho_{\mathrm{i}}$. The energy flux at
$\ell$ is given by
\begin{equation}
 \varepsilon \sim 
  \frac{n_{0\mathrm{i}}m_{\mathrm{i}}v_{\mathrm{th,i}}^{2}}{4\tau_{\ell}}
  \left(\frac{h_{\mathrm{i},\ell}v_{\mathrm{th,i}}^{3}}{n_{0\mathrm{i}}}\right)^{2}= \textrm{const}.
\end{equation}
From the gyrokinetic equation, the nonlinear decorrelation time is estimated as
\begin{equation}
 \frac{1}{\tau_{\ell}} \sim
  \frac{1}{B_{0}} \frac{1}{\ell^{2}}
  \left\langle \phi \right\rangle_{\bm{R}}
  \sim \frac{1}{B_{0}} \frac{\phi_{\ell}}{\ell^{2}}
  \left(\frac{\rho_{\mathrm{i}}}{\ell}\right)^{-1/2}.
\end{equation}
The gyro-averaging operation introduces a reduction factor
$\left(\rho_{\mathrm{i}}/\ell\right)^{-1/2}$. Note that
$A_{\parallel}\approx0$. From the quasi-neutrality for the Boltzmann response electron,
\begin{equation}
 \frac{q_{\mathrm{i}}\phi_{\ell}}{T_{0\mathrm{i}}}
  \left(1+Q\right)
  \sim \frac{v_{\mathrm{th,i}}^{3}}{n_{0\mathrm{i}}} h_{\mathrm{i},\ell}
  \left( \frac{\rho_{\mathrm{i}}}{\ell} \right)^{-1/2}
  \left( \frac{\delta v}{v_{\mathrm{th,i}}}\right)^{1/2}.
  \label{eq:pmqn_scale}
\end{equation}
The nonlinear phase mixing produces structures in velocity space
correlated with the spatial scale,
\begin{equation}
 \frac{\delta v}{v_{\mathrm{th,i}}} \sim \frac{\ell}{\rho_{\mathrm{i}}}.
  \label{eq:pmperp_scale}
\end{equation}
Combining all the relations, we \revproof{obtain}
\begin{align}
 \frac{\varepsilon}{n_{0\mathrm{i}}m_{\mathrm{i}}} & \sim
 \frac{v_{\mathrm{th,i}}^{12}}{8n_{0\mathrm{i}}^{3}}
 \left(1+Q\right)^{-1} \ell^{-1/2}\rho_{\mathrm{i}}^{-1/2}
 h_{\mathrm{i},\ell}^{3}.
 \end{align}
 This gives a scaling for $h_{\mathrm{i},\ell}$ as
 \begin{equation}
  h_{\mathrm{i},\ell} \sim 
   \frac{n_{0\mathrm{i}}}{v_{\mathrm{th,i}}^{4}}
   \left(1+Q\right)^{1/3}
   \left(\frac{\varepsilon}{n_{0\mathrm{i}}m_{\mathrm{i}}}\right)^{1/3}
   \ell^{1/6}\rho_{\mathrm{i}}^{1/6}.
  \label{eq:entscale_h}
 \end{equation}
 Scalings for $\phi_{\ell}$ and $\tau_{\ell}$ are similarly given by
 \begin{align}
  \revproof{\frac{\phi_{\ell}}{B_{0}}} & \sim
  \left(1+Q\right)^{-2/3}
  \left(\frac{\varepsilon}{n_{0\mathrm{i}}m_{\mathrm{i}}}\right)^{1/3}
  \ell^{7/6} \rho_{\mathrm{i}}^{1/6},
  \label{eq:entscale_phi}
  \\
  \tau_{\ell} & \sim
  \left(1+Q\right)^{2/3}
  \left(\frac{\varepsilon}{n_{0\mathrm{i}}m_{\mathrm{i}}}\right)^{-1/3}
  \ell^{1/3} \rho_{\mathrm{i}}^{1/3}.
  \label{eq:entscale_tau}
 \end{align}
The energy scalings follow from these as
 \begin{align}
  W_{h,k} \diff k & = \int
  \frac{T_{0\mathrm{i}}\left|h_{\mathrm{i},k}\right|^{2}}{2f_{0\mathrm{i}}} \diff \bm{v}
  \sim n_{0\mathrm{i}} m_{\mathrm{i}} 
  \left(1+Q\right)^{2/3}
  \left(\frac{\varepsilon}{n_{0\mathrm{i}}m_{\mathrm{i}}}\right)^{2/3}
  k^{-1/3} \rho_{\mathrm{i}}^{1/3},
  \label{eq:esgk_h_spec}
  \\
  W_{\phi,k} \diff k & =
  \frac{q_{\mathrm{i}}^{2} n_{0\mathrm{i}}}{2T_{0\mathrm{i}}}
  \left|\phi_{k}\right|^{2}
  \sim n_{0\mathrm{i}} m_{\mathrm{i}}
  \left(1+Q\right)^{-4/3}
  \left(\frac{\varepsilon}{n_{0\mathrm{i}}m_{\mathrm{i}}}\right)^{2/3}
  k^{-7/3} \rho_{\mathrm{i}}^{-5/3}.
  \label{eq:esgk_phi_spec}
 \end{align}

The collisional cutoff scale in Fourier space $k_{\mathrm{d}}$ and in
velocity space $\delta v_{\mathrm{d}}$ is estimated by balancing the
cascade time and collisional time scale,
\begin{equation}
 \nu_{\mathrm{ii}} v_{\mathrm{th,i}}^{2} \left(\pdf{}{v}\right)^{2} \sim \tau_{\ell}^{-1}.
\end{equation}
Substituting the relations \eqref{eq:pmperp_scale} \revproof{and}
\eqref{eq:entscale_tau} at $\ell^{-1}=k_{\mathrm{d}}$, we obtain
\begin{equation}
 k_{\mathrm{d}}\rho_{\mathrm{i}} \sim
  \left(\nu_{\mathrm{ii}}\tau_{\rho_{\mathrm{i}}}\right)^{-3/5}
  \sim D^{3/5},
\end{equation}
where the dimensionless number defined by
\begin{equation}
 D=\left(\nu_{\mathrm{ii}}\tau_{\rho_{\mathrm{i}}}\right)^{-1},
\end{equation}
is called the Dorland number characterizing the range of ion kinetic
scale in velocity space as well as in Fourier space. Setting
$\ell=\rho_{\mathrm{i}}$ in \eqref{eq:entscale_tau}, we obtain
\begin{equation}
 D \sim \nu_{\mathrm{ii}}^{-1} (1+Q)^{-2/3}
  \left( \frac{\varepsilon}{n_{0\mathrm{i}}m_{\mathrm{i}}}\right)^{1/3}
  \rho_{\mathrm{i}}^{-2/3}.
\end{equation}

\begin{figure}
 \centering
 \includegraphics[scale=0.8]{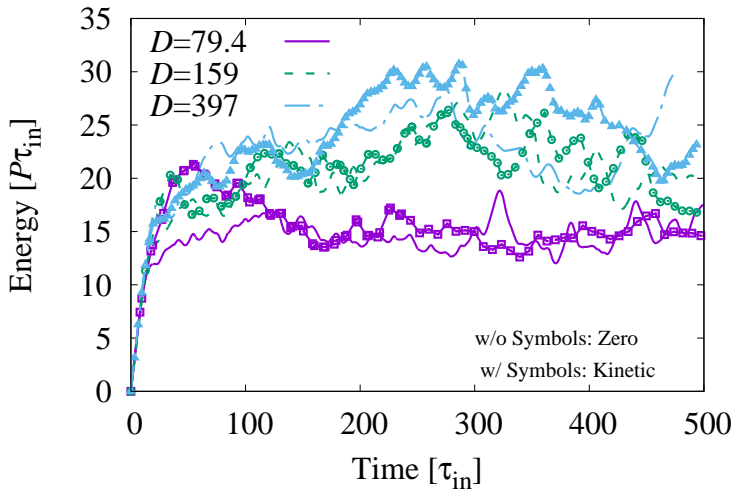}
 \caption{\label{fig:ioncasc_tevo_ene}Time evolution of the energy for the
 $k_{\mathrm{in}}\rho_{\mathrm{i}}=2=O(1)$ case. Three cases
 of $D$ values with two electron models are simulated. There is not
 much difference between zero response and gyrokinetic electrons.}
\end{figure}

\begin{figure}
 \centering
 \includegraphics[scale=0.8]{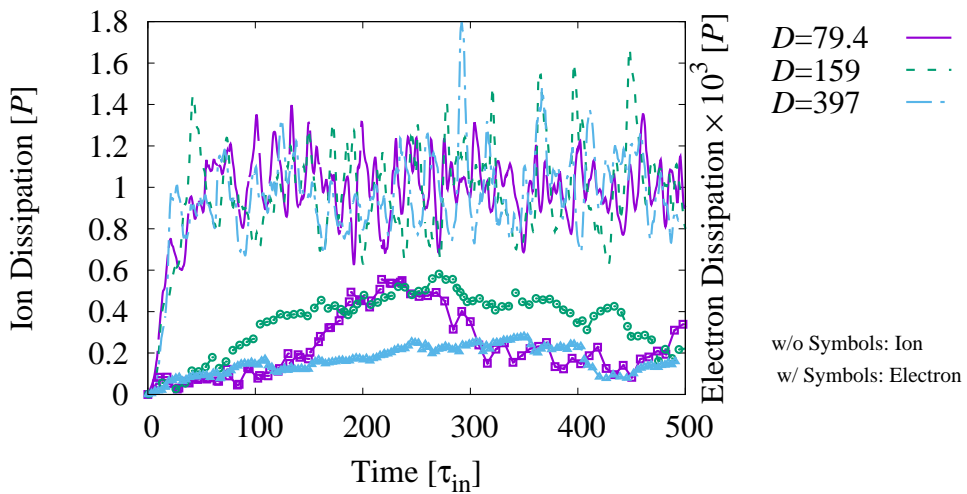}
 \caption{\label{fig:ioncasc_tevo_diss}Time evolution of the dissipation
 for the $k_{\mathrm{in}}\rho_{\mathrm{i}}=2=O(1)$ case. The ion 
 dissipation behaves similarly for both electron response
 cases, so only the gyrokinetic electron case is shown. The electron
 dissipation, which exists only in the gyrokinetic case, is much smaller
 than that of ions. \revproof{Note} that the electron dissipation is magnified by a
 factor of $10^{3}$.}
\end{figure}

\begin{figure}
 \centering
 \includegraphics[scale=0.8]{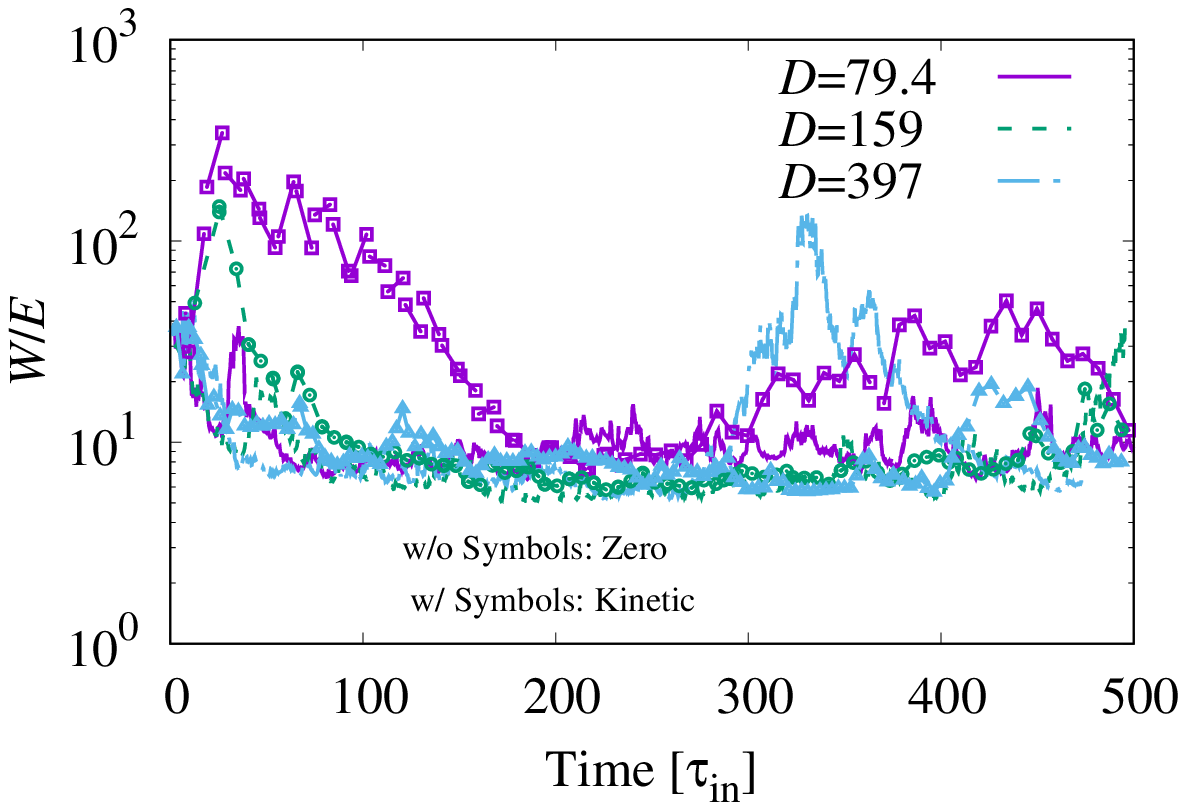}
 \caption{\label{fig:ioncasc_tevo_wovere}The ratio of the generalized
 energy to the electrostatic invariant $W/E$.}
\end{figure}

To study turbulence in the ion kinetic regime, we take the injection
scale at the ion Larmor radius $k_{\mathrm{in}}\rho_{\mathrm{i}}=2$. To
reduce the effect of the inverse cascade, we set the box size as
$k_{\mathrm{in}}L=1$. Other parameters are the same as those in the previous
section. We perform simulations for three values of $\nu_{\mathrm{ii}}$
corresponding to (i) $D=79.4$, (ii) $D=159$ \revproof{and} (iii) $D=397$ to vary
$k_{\mathrm{d}}\rho_{\mathrm{i}}$.
In the estimation of $D$, we set $Q=0$ and
$\varepsilon=\overline{P}_{\mathrm{in}}$ for simplicity. The gyrokinetic
and zero response electron models are used. For all cases, the
resolution in the $x$--$y$ plane is $128^{2}$ and that in velocity space is
$64^{2}$ for \revproof{cases} (i) \revproof{and} (ii) and $128^{2}$ for
\revproof{case} (iii). The time evolution of the generalized energy in
\revproof{figure}~\ref{fig:ioncasc_tevo_ene} confirms that the system reaches steady
states for all cases. In this ion kinetic scale, the ion entropy cascade
occurs in velocity space until $\delta v$ extends to the scale
where the collisional dissipation of ions (proportional to
$\nu_{\mathrm{ii}} \delta v^{2}$) 
balances with the input, namely the injected energy is completely taken
out from the system at the ion scale. As long as the sufficiently large resolution is
taken in both real and velocity spaces, the steady state is achieved. We
see that there is not much difference between the zero response and
gyrokinetic electron cases. For the gyrokinetic case, the electron
dissipation is much smaller than that of ions
(\revproof{figure}~\ref{fig:ioncasc_tevo_diss}). \revise{In
\revproof{figure}~\ref{fig:ioncasc_tevo_wovere}, we plot the ratio of
the generalized energy to the electrostatic invariant, $W/E$. The ratio
seems approaching to some constant value ($\sim10$) regardless of the
parameters. This fact may have some implications to the nature of
inverse cascades, which will be studied elsewhere.}


\begin{figure}
 \centering
 \includegraphics[scale=0.5]{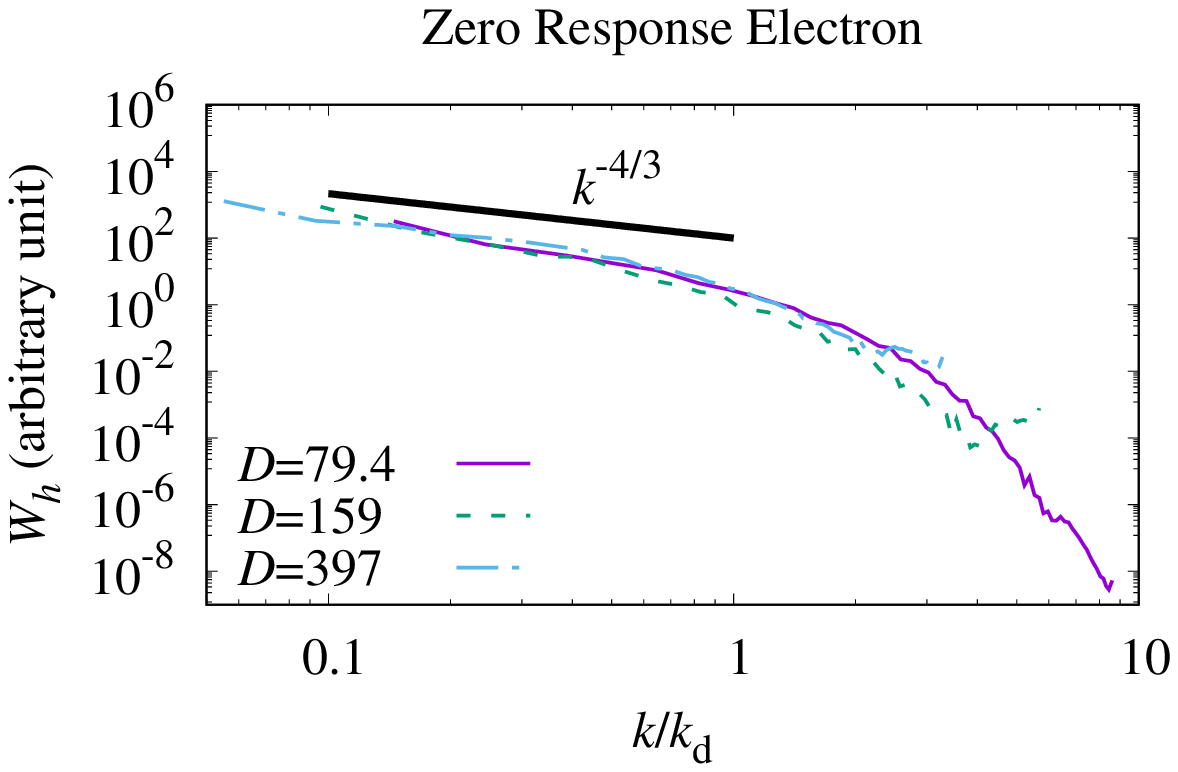}
 \includegraphics[scale=0.5]{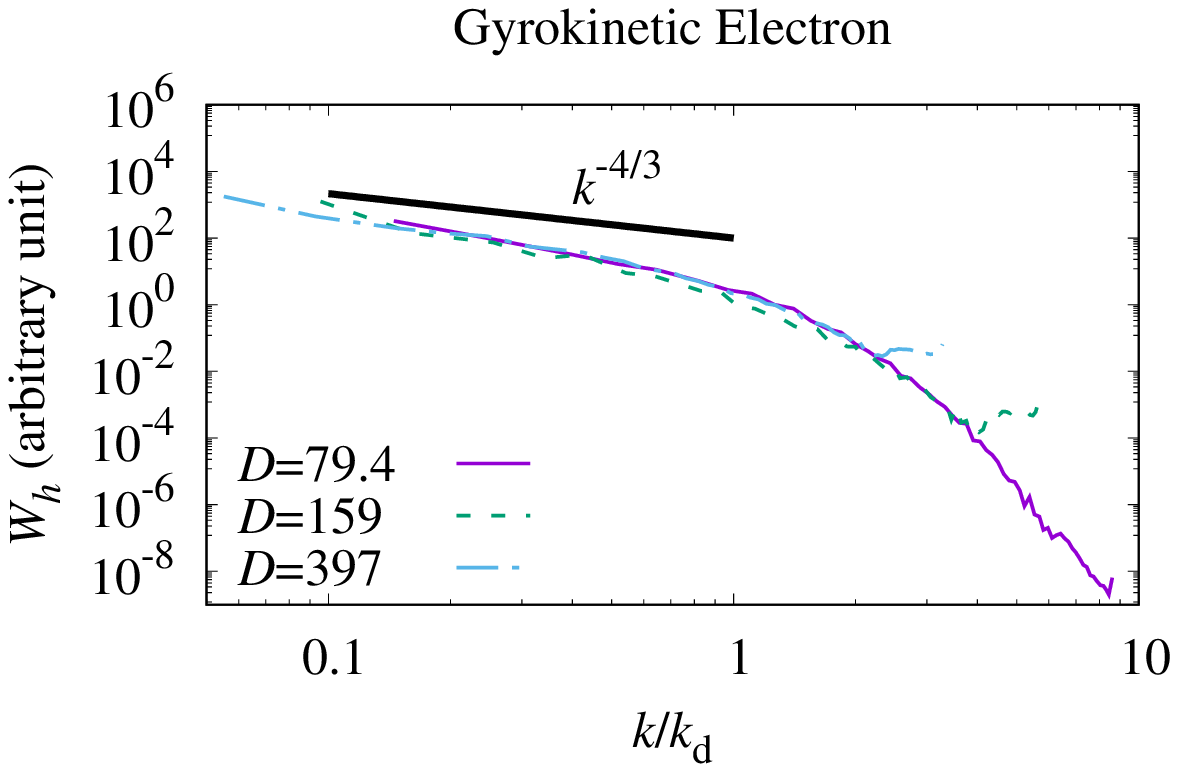}
 \includegraphics[scale=0.5]{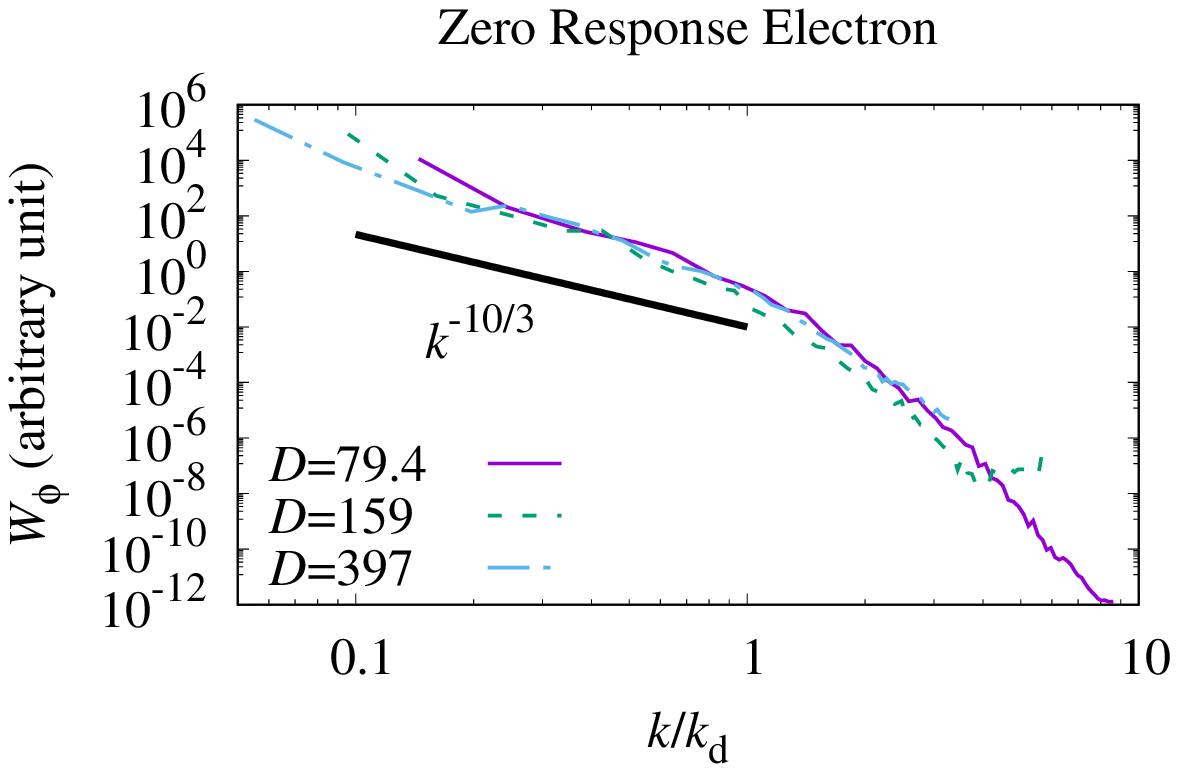}
 \includegraphics[scale=0.5]{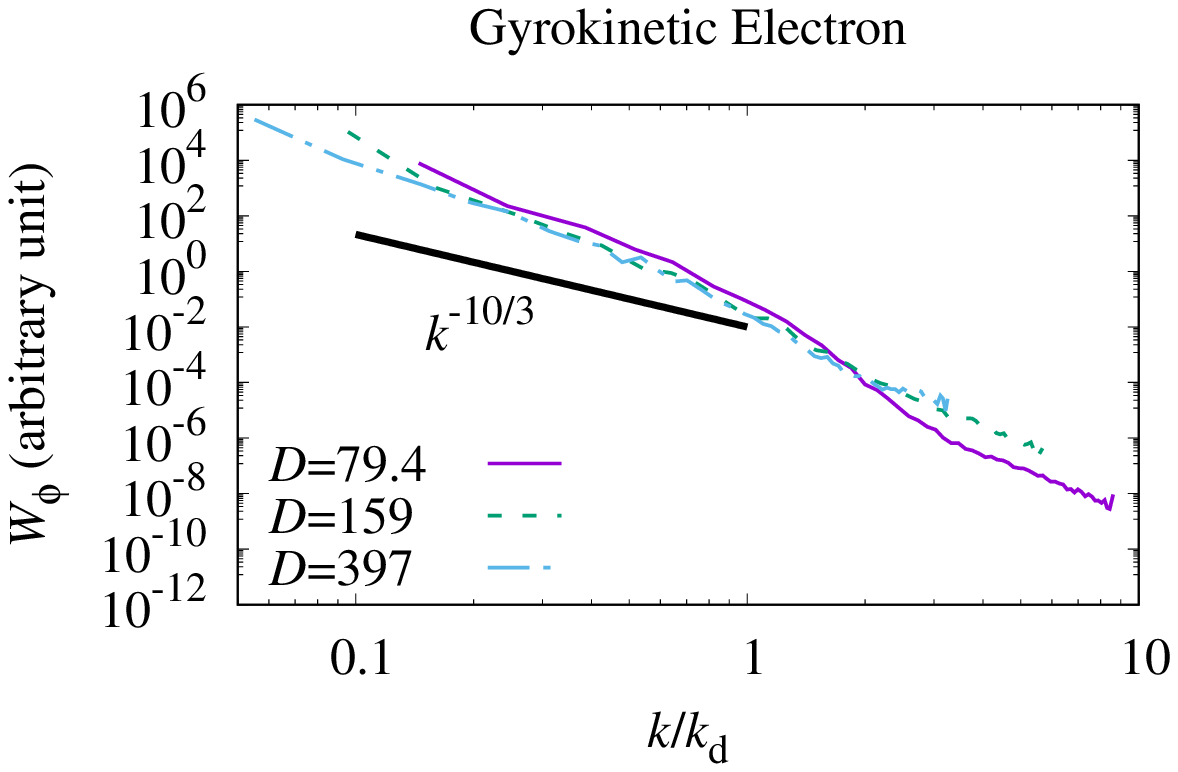}
 \caption{\label{fig:ioncasc_spec}Wavenumber spectra of
 $\overline{W}_{h}$ and $\overline{W}_{\phi}$. The zero response and
 gyrokinetic electron models are shown.}
\end{figure}

Figure~\ref{fig:ioncasc_spec} shows the wavenumber spectra of $\overline{W}_{h}$
and $\overline{W}_{\phi}$. Although both electron response cases are shown in
separate plots, we again \revproof{note} that the spectra of the ion entropy
cascade (as well as the temporal evolution of the energy) are
indifferent to the electron response.
The wavenumber is normalized by the dissipation cutoff scale
$k_{\mathrm{d}}$. For all cases, the spectra start to fall off around
$k\sim k_{\mathrm{d}}$. \revproof{Here} $k_{\mathrm{d}}$ gives a good
measure of the dissipation scale. For the largest $D$, 
$k_{\mathrm{d}}\rho_{\mathrm{i}}\sim36.2$ and the range of injection and
dissipation are separated. We see spectra roughly consistent with the
theory in the range $k/k_{\mathrm{d}}<0.5$ although the dynamic range is
quite narrow.

We do not reproduce all the detail of the entropy cascade dynamics, such
as velocity space spectra, entropy transfer and just remark that they are
almost the same as those for the well-established decaying turbulence
case. Interested readers are referred to the work by \citet{TatsunoDorlandSchekochihin_09,TatsunoBarnesCowley_10,TatsunoPlunkBarnes_12}. The
virtue of the driven case developed here is that simulations are set up
in a controlled manner where the injection and dissipation scales are
fixed, and long-time statistics can be discussed in a steady state if
necessary.

\subsection{Compressive fluctuations in high-$\beta$ plasmas}

In high-$\beta$ cases, compressive fluctuations involving $\delta
B_{\parallel}$ are also induced. The parallel magnetic fluctuation is
given by
\begin{equation}
 \frac{\delta B_{\parallel,\bm{k}}}{B_{0}} = - \frac{\beta_{\mathrm{i}}}{2}
  \frac{1}{n_{0\mathrm{i}}}\int \frac{2v_{\perp}^{2}}{v_{\mathrm{th,i}}^{2}} 
  \frac{\revproof{J_{1\mathrm{i}}}}{\alpha_{\mathrm{i}}} h_{\mathrm{i},\bm{k}}\diff \bm{v},
\end{equation}
if electrons follow the Boltzmann relation ($h_{\mathrm{e}}=0$). The
magnetic fluctuations are
small compared with $q_{\mathrm{i}} \phi/T_{0\mathrm{i}}$ by a factor of
$\beta_{\mathrm{i}}/(k\rho_{\mathrm{i}})$. By employing a similar
argument, we obtain a scaling law for the magnetic fluctuation, 
\begin{align}
 \frac{\delta B_{\parallel,\ell}}{B_{0}}
 & \sim \beta_{\mathrm{i}}
 \frac{1}{v_{\mathrm{th,i}}}
 \left(1+Q\right)^{1/3}
 \left( \frac{\varepsilon}{n_{0\mathrm{i}}m_{\mathrm{i}}}
 \right)^{1/3}
 \ell^{13/6} \rho_{\mathrm{i}}^{-11/6},
\\
 \overline{M}_{\parallel,k} \diff k
 = \frac{\left|\delta B_{\parallel,k}\right|^{2}}{2\mu_{0}}
 & \sim
 n_{0\mathrm{i}} m_{\mathrm{i}} \beta_{\mathrm{i}} 
 \left( 1 + Q \right)^{2/3}
 \left( \frac{\varepsilon}{n_{0\mathrm{i}}m_{\mathrm{i}}}\right)^{2/3}
 k^{-13/3} \rho_{\mathrm{i}}^{-11/3}.
 \label{eq:magspec}
\end{align}

\begin{figure}
 \centering
 \includegraphics[scale=0.8]{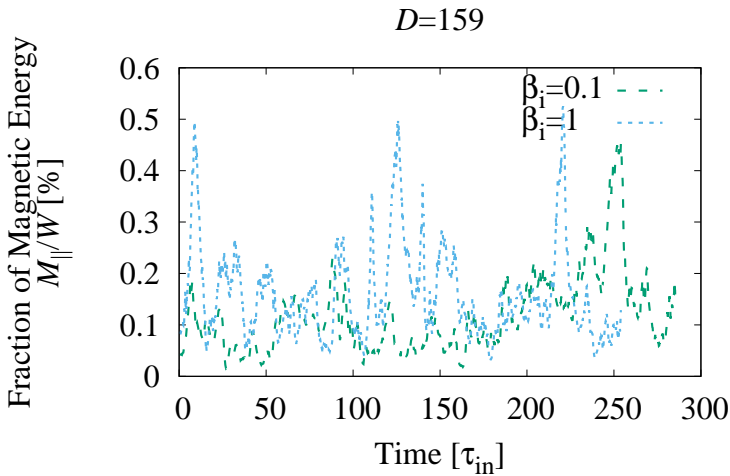}
 \caption{\label{fig:ioncasc_tevo_ene_bpar}Time evolution of the
 fraction of magnetic energy to total energy $M_{\parallel}/W$.}
\end{figure}

\begin{figure}
 \centering
 \includegraphics[scale=0.8]{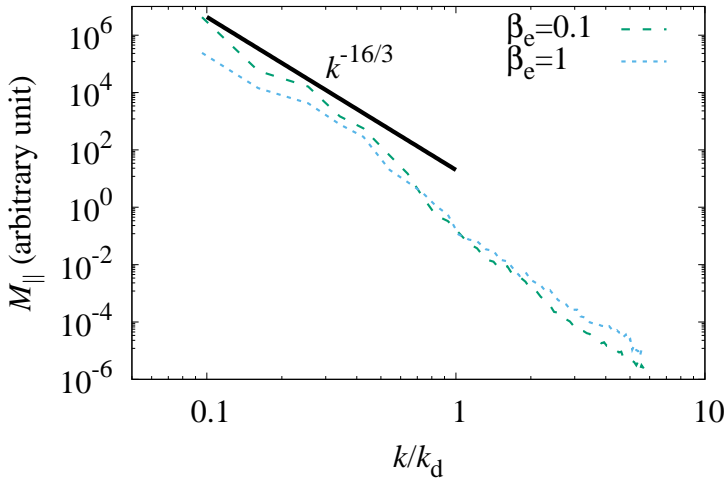}
 \caption{\label{fig:ioncasc_spec_bpar}Wavenumber spectra of $M_{\parallel}$.}
\end{figure}

We include finite $\beta_{\mathrm{i}}=0.1, 1$ for the $D=159$ case, and
observe the spectrum of
$\overline{M}_{\parallel}$. Figure~\ref{fig:ioncasc_tevo_ene_bpar} shows
the fraction of parallel magnetic energy to the total energy. The magnetic
energy stays at a negligible level throughout the simulation. The
magnetic energy spectrum shown in \revproof{figure}~\ref{fig:ioncasc_spec_bpar}
agrees well with the theoretical predicted scaling $k^{-16/3}$ for any
values of $\beta_{\mathrm{i}}$.


\revproof{As} the compressive fluctuations are small, they do not affect the
flow dynamics, and behave as a passive scalar even though they are
essentially coupled with the flows in this kinetic regime. When $\beta$ gets
even higher, the magnetic fluctuations may become comparable to other
components and start to affect the flows.
This result may be compared with the compressive fluctuations
in Alfv\'enic turbulence observed in solar winds and simulations of the
electromagnetic plasma turbulence. The long-standing puzzle of the
magnetic fluctuations in solar winds~\citep{MarschTu_90,Chen_16} is that
they do not damp and have the scaling similar to the fluid. It is argued
that the stochastic plasma echo inhibits phase mixing so that plasmas
behave more fluid-like
(fluidisation)~\citep{SchekochihinParkerHighcock_16,MeyrandKanekarDorland_19},
thus follow shallower spectra than the kinetic one
\eqref{eq:magspec}. The present result showing the kinetic
nature of turbulence shows a marked difference with the Alfv\'enic
case. We remark that here the Landau damping mechanism of compressive
fluctuations along the mean magnetic field is absent.

\section{Summary}
\label{sec:summary}

We have constructed an external forcing term to drive turbulence in
two-dimensional electromagnetic gyrokinetics and have implemented it in
{\tt AstroGK} code. The additional term in the gyrokinetic equation
describing the temporal evolution of distribution functions induces density
and temperature fluctuations of plasmas, and thus disturbs the electrostatic
potential and magnetic field parallel to the mean magnetic field
through the quasi-neutrality and pressure balance equations. The induced
electrostatic potential corresponds to the $\EB$ velocity, therefore the
term is equivalent to the \revproof{`forcing'} (drives flows) in the reduced MHD
equations. By construction, the newly introduced term does not excite
the magnetic field perpendicular to the mean magnetic field. The
Alfv\'en wave is completely absent in the present work.

By the external drive, stationary turbulence is numerically
generated in a controlled manner to study \revise{the} statistical properties
of fluctuations. The two essential properties of the forcing
are locality in the wavenumber space and a pre-determined energy
injection rate. To fix an energy injection rate, 
the force is chosen to \revproof{make the force--velocity correlation
vanish}.
These properties enable \revproof{turbulence to be set up} in any
dynamical scales ranging from the injection
scale $k_{\mathrm{in}}$ set by the forcing term to the dissipation scale
$k_{\mathrm{d}}$ determined by the injected power and dissipation
parameter (collisionality). We have performed two simulations,
namely \revproof{in the large-scale} fluid regime and
\revproof{small-scale} ion kinetic regime.

In the fluid regime ($k_{\mathrm{in}} \rho_{\mathrm{i}}\ll1$), we have
shown the forward enstrophy and inverse energy cascade by gyrokinetic
simulations, and have successfully produced the theoretical scalings of
the wavenumber spectra. We have also demonstrated the breaking of the
wavenumber spectra due to the ion polarisation effect, corresponding to
the \revproof{CHM} turbulence. These results are all well studied
using fluid models, and there is no reason to employ computationally
demanding kinetic description. However, we emphasize that the ability to
correctly capture \revproof{large-scale} behaviour using the kinetic model is
still useful for studying the connection of fluid models and kinetic
models, and also for performing holistic simulations including
all microphysics to \revproof{macro-scale} dynamics. We note that in
the fluid regime, it is a subtle issue to achieve a stationary state
because ignoring small scales prevents collisional dissipation from
developing to balance the injection.

In the kinetic regime ($k_{\mathrm{in}} \rho_{\mathrm{i}} \sim 1$), we
have studied the ion entropy cascade turbulence. In this regime, the
nonlinear phase mixing due to the finite Larmor radius effect creates
structures in velocity space. The ion entropy cascade carries the
injected energy to small scales in both real and velocity spaces, and
eventually ceases at the dissipation cutoff scale due to
collisions. It is found that the ion collisional dissipation
balances with the injected energy in this regime. The obtained
scalings agree well with the theoretical prediction and with decaying turbulence
simulations which were previously done. In addition to the electrostatic
case, we have also performed simulations of high-$\beta$ plasmas where
compressive fluctuations involving magnetic \revise{fields} parallel to the mean
field are also excited. It is confirmed that the compressive fluctuations are
just passively advected by the flow, and do not affect the scaling laws
of turbulence as long as $\beta$ is not too high.

The simulations performed in this work are in
\revproof{four-dimensional} phase space
(two each in real and velocity spaces), and the resolution is up to
$O(10^2)$ in each direction, which is relatively low compared with
the cutting edge numerical simulations. We admit the dynamic range is
not sufficiently wide to observe clear spectra because of the
restriction of numerical resources, although we do capture the essential
features of generated turbulence. It is acceptable for the testing
purpose of the development of the forcing method.

The turbulence generated by the forcing method presented here is
restricted to two dimensions and the case without Alfv\'en
waves on purpose. Such a situation may occur in realistic plasmas
because there are several energy cascade channels from the macroscopic to kinetic
scales, and it is not known on which paths the cascades take place depending
on situations. The current study focusing on a specific situation
still contributes to the comprehensive understanding of kinetic plasma
turbulence. In the future, we apply this work for the situation in the
presence of an initial magnetic field and magnetic reconnection, which
provides another energy dissipation mechanism. \revproof{As} the relative
strength of turbulence compared with magnetic reconnection is
measurable, the developed method enables quantitative analyses of energy
dissipation in the co-existing system of turbulence and magnetic reconnection.


\section*{Acknowledgements}

The author would like to thank W. Dorland\revise{, Y. Kawazura}, N.~F. Loureiro,
G. Plunk and T. Tatsuno for \revise{their} fruitful comments.
The computation of this work is performed on the facilities of the
Center for Cooperative Work on Computational Science, University of Hyogo, 
the JFRS-1 supercomputer system at \revproof{the} Computational Simulation Centre of
International Fusion Energy Research Centre (IFERC-CSC) in Rokkasho
Fusion Institute of QST (Aomori, Japan) and on \revproof{`Plasma Simulator'} of
NIFS with the support and under the auspices of the NIFS Collaboration
Research program (NIFS18KNSS112).

\section*{Declaration of interests}

The author reports no conflict of interests.

\pagebreak
\appendix
\section{Gyrokinetic-Maxwell equations}
\label{sec:gkm}

In this paper, we implement an additional forcing term in {\tt
AstroGK}. We briefly describe the gyrokinetic-Maxwell equations used in
the code. (See \citet{NumataHowesTatsuno_10} for the complete description of the
model.) {\tt AstroGK} solves the nonlinear, local flux tube, $\delta f$, 
electromagnetic gyrokinetic equation. It determines fluctuating fields
of the species $s$ around a Maxwellian background in the mean magnetic field
$\bm{B}_{0}=B_{0}\bm{z}$ defined by $n_{0s}$ and $T_{0s}$:
$f_{0s}=n_{0s}/(\sqrt{\upi}v_{\mathrm{th},s})^{3} \exp
(-v^2/v_{\mathrm{th},s}^{2})$ with
$v_{\mathrm{th},s}=\sqrt{2T_{0s}/m_{s}}$ being the thermal velocity.
For the sake of simplicity, we only consider \revproof{two-dimensional}
($\p/\p_{z}=0$) and uniform plasmas.

The gyrokinetic equation determines the evolution of $g_{s}$ in Fourier space
\begin{equation}
 \pdf{g_{\bm{k},s}}{t}
 + \frac{1}{B_{0}}
 {\mathcal F}\left( \left\{ \langle \chi \rangle_{\bm{R}_{s}}, h_{s}
 \right\}\right)
 = \frac{q_{s} f_{0s}}{T_{0s}}
 v_{\parallel} J_{0s} \pdf{A_{\parallel,\bm{k}}}{t}
 + C_{\bm{k}} (h_{\bm{k},s}),
\end{equation}
where $g_{s}$ is defined using the non-Boltzmann part of the perturbed
distribution function $h_{s} = \delta f_{s} + (q_{s} \phi/T_{0s}) f_{0s}$ as
\begin{equation}
 g_{\bm{k},s} = h_{\bm{k},s}
  - \frac{q_{s} \phi_{\bm{k}}}{T_{0s}}
 J_{0s} f_{0s}
 - \frac{2v_{\perp}^{2}}{v_{\mathrm{th},s}^{2}}
 \frac{J_{1s}}{\alpha_{s}} f_{0s}
 \frac{\delta B_{\parallel,\bm{k}}}{B_{0}}.
\end{equation}
Note that $\bm{k}=(k_{x},k_{y})$ is the wavenumber in the plane
perpendicular to $\bm{B}_{0}$. The distribution function $g_{s}$ and
field variables (e.g. $\phi$) are decomposed, respectively, in the
\revproof{gyro-centre} coordinate $\bm{R}_{s}$ and particle coordinate $\bm{r}$ as
\begin{align}
 g_{s} & = \sum_{\bm{k}} g_{\bm{k},s} e^{\imag \bm{k}\cdot\bm{R}_{s}}, \\
 \phi & = \sum_{\bm{k}} \phi_{\bm{k}} e^{\imag \bm{k}\cdot\bm{r}}.
\end{align}
The coordinate transform is given by
\begin{equation}
 \bm{R}_{s} = \bm{r} -
 \frac{\bm{v}\times\bm{B}_{0}}{\Omega_{s}}
~~~ (\Omega_{s}\textrm{: cyclotron frequency}).
\end{equation}
The Bessel function of the first kind $J_{ns}=J_{n}(\alpha_s)$ with the
argument $\alpha_{s}=|\bm{k}|v_{\perp}/\Omega_{s}$ arises from the
coordinate transformation via the gyro-averaging operation.

Three field variables, $\phi$, $A_{\parallel}$ \revproof{and} $\delta B_{\parallel}$,
satisfy the Maxwell equations:
\begin{align}
 \left[ \sum_{s} \frac{q_{s}^{2}n_{0s}}{T_{0s}}\left(1-\Gamma_{0s}\right) \right]
 \phi_{\bm{k}}
 - \left[ \sum_{s} q_{s} n_{0s} \Gamma_{1s} \right]
 \frac{\delta B_{\parallel,\bm{k}}}{B_{0}}
 & = \sum_{s} q_{s} \int \revproof{J_{0s}} g_{\bm{k},s} \diff \bm{v},
 \label{eq:gkg_f_qn} \\
 k^{2} A_{\parallel,\bm{k}} & = \mu_{0} \sum_{s} q_{s}
 \int v_{\parallel} \revproof{J_{0s}} g_{\bm{k},s} \diff \bm{v},
 \label{eq:gkg_f_ampere_para} \\
 \left[ \sum_{s} q_{s}n_{0s} \Gamma_{1s} \right]
 \phi_{\bm{k}}
 + \left[ \frac{B_{0}^{2}}{\mu_{0}} + \sum_{s} n_{0s} T_{0s} \Gamma_{2s} \right]
 \frac{\delta B_{\parallel,\bm{k}}}{B_{0}}
 & = - \sum_{s} 
\int m_{s} v_{\perp}^{2} \frac{\revproof{J_{1s}}}{\alpha_{s}} g_{\bm{k},s}
 \diff \bm{v}.
 \label{eq:gkg_f_ampere_perp}
\end{align}
The function $\Gamma_{ns}=\Gamma_{n}(b_{s})$ is given by
\begin{align}
 \Gamma_{0s} & = I_{0}(b_{s}) e^{-b_{s}}, \\
 \Gamma_{1s} & = \left(I_{0}(b_{s}) - \revproof{I_{1}}(b_{s})\right) e^{-b_{s}}, \\
 \Gamma_{2s} & = 2 \Gamma_{1s},
\end{align}
where $I_{n}$ is the modified Bessel function of the first kind, and the
argument is $b_{s}=(|\bm{k}|\rho_{s})^{2}/2$ ($\rho_s$ is the Larmor radius).

We introduce the following symbols for notational simplicity.
\begin{equation}
 \revproof{{\mathsfbi Y}} = 
 \begin{pmatrix}
  \revproof{{\mathsfit Y}}_{1} &
  - \revproof{{\mathsfit Y}}_{3} \\ 
  \revproof{{\mathsfit Y}}_{3} &
  \revproof{{\mathsfit Y}}_{2}
 \end{pmatrix}
 =
 \begin{pmatrix}
  \sum_{s} \frac{q_{s}^{2}n_{0s}}{T_{0s}} \left(1-\Gamma_{0s}\right)
  &
  - \sum_{s} q_{s} n_{0s} \Gamma_{1s}
  \\
  \sum_{s} q_{s} n_{0s} \Gamma_{1s}
  & 
  \frac{B_{0}^{2}}{\mu_{0}} + \sum_{s} n_{0s} T_{0s} \Gamma_{2s}
 \end{pmatrix}.
\end{equation}
Then, the coupled field equations for $\phi$ and $\delta B_{\parallel}$
are given by
\begin{equation}
 \begin{pmatrix}
  \revproof{{\mathsfit Y}}_{1} & - \revproof{{\mathsfit Y}}_{3} \\
  \revproof{{\mathsfit Y}}_{3} & \revproof{{\mathsfit Y}}_{2}
 \end{pmatrix}
 \begin{pmatrix}
  \phi_{\bm{k}} \\
  \frac{\delta B_{\parallel,\bm{k}}}{B_{0}}
 \end{pmatrix}
 =
 \begin{pmatrix}
 \sum_{s} q_{s} \int \revproof{J_{0s}} g_{\bm{k},s} \diff \bm{v} \\
  - \sum_{s} q_{s} \int m_{s} v_{\perp}^{2}
  \frac{\revproof{J_{1s}}}{\alpha_{s}}
  g_{\bm{k},s} \diff \bm{v}
 \end{pmatrix}.
\end{equation}
In some cases, we do not solve the gyrokinetic equation for electrons
and just assume the zero response ($\delta f_{\mathrm{s}}=0$ ) or the
adiabatic (or Boltzmann) response ($h_{\mathrm{e}}=0$). In either case, the electron
response is parameterized by $Q_{\mathrm{e}}$, where
\begin{equation}
 \revproof{{\mathsfbi Y}} =
 \begin{pmatrix}
  Q_{\mathrm{e}} + \frac{q_{\mathrm{i}}^{2}
  n_{0\mathrm{i}}}{T_{0\mathrm{i}}}
  \left(1-\Gamma_{0\mathrm{i}}\right)
  & 
  - q_{\mathrm{i}} n_{0\mathrm{i}} \Gamma_{0\mathrm{i}}
  \\
  q_{\mathrm{i}} n_{0\mathrm{i}} \Gamma_{0\mathrm{i}}
  &
  \frac{B_{0}^{2}}{\mu_{0}} + n_{0\mathrm{i}} T_{0\mathrm{i}} \Gamma_{2\mathrm{i}}
 \end{pmatrix}.
\end{equation}
For the zero response case, $Q_{\mathrm{e}}=0$, \revproof{whereas} for the adiabatic
response case, $Q_{\mathrm{e}}=q_{\mathrm{e}}^{2} n_{0\mathrm{e}}/T_{0\mathrm{e}}$.
The right-hand side of the field equations should also change
accordingly ($g_{\bm{k},\mathrm{e}}=0$).

Other non-standard symbols that are not explained in the text are
summarized in table~\ref{tbl:gk_symbol}.
\begin{table}
 \centering
 \caption{\label{tbl:gk_symbol}Explanation of non-standard symbols.}
 \begin{tabular}{cl}
  ${\mathcal F}$ & Fourier transform \\
  $\left\{a,b\right\}\equiv(\p_{x}a)(\p_{y}b)-(\p_y a)(\p_x b)$ & Poisson bracket \\
  $\langle \dots \rangle_{\bm{R}_s}$ & gyro-average \\
  $C_{\bm{k}}$ & collision operator at the gyro-center coordinate $\bm{R}_{s}$\\
  $\perp$, $\parallel$ & directions perpendicular/parallel to $\bm{B}_{0}$
 \end{tabular}
\end{table}


\bibliographystyle{jpp}
\bibliography{gk_drive}

\end{document}